\documentclass[journal,transmag]{IEEEtran}

\usepackage[nocompress]{cite}
\usepackage{enumerate}
\usepackage{amssymb}
\usepackage{multirow}
\usepackage{amsmath}
\usepackage[english]{babel}
\usepackage{graphicx}
\usepackage{stfloats}
\usepackage{color}
\usepackage{amssymb}
\usepackage{amsthm}
\usepackage{bm}
\usepackage{algorithm}
\usepackage{algorithmic}

\newtheorem{myDef}{Definition}
\newtheorem{myThe}{Theorem}

\newtheorem{myLem}{Lemma}

\begin{document}
\title{Game-Theoretic Design of Optimal Two-Sided Rating Protocols for Service Exchange Dilemma in Crowdsourcing}

\author{Jianfeng Lu,~\IEEEmembership{Member,~IEEE,}
        Yun Xin,~\IEEEmembership{}
        Zhao Zhang,~\IEEEmembership{Member,~IEEE,}
        Xinwang Liu,~\IEEEmembership{Member,~IEEE,}\\
        Kenli Li~\IEEEmembership{Senior Member,~IEEE}


\IEEEcompsocitemizethanks{\IEEEcompsocthanksitem Jianfeng Lu is with the Department of Computer Science and Engineering, Zhejiang Normal University, Jinhua, Zhejiang, China, and also with State Key Laboratory for Novel Software Technology, Nanjing University, Nanjing, Jiangsu, China (e-mail: lujianfeng@zjnu.cn)
\IEEEcompsocthanksitem Yun Xin, Zhao Zhang and Shasha Yang are with the Department of Computer Science and Engineering, Zhejiang Normal University, Jinhua, Zhejiang, China.(e-mail: xinyun\_zjnu@163.com; zhaozhang@zjnu.cn)
\IEEEcompsocthanksitem Xinwang Liu is with the school of of Computer, National University of Defense Technology, Changsha, China.(e-mail: xinwangliu@nudt.edu.cn)
\IEEEcompsocthanksitem Kenli Li is with the College of Computer Science and Electronic Engineering, Hunan University, Changsha, Hunan, China, and the National Supercomputing Center in Changsha, Changsha, Hunan 410082, China.(e-mail:lkl@hnu.edu.cn)
}
\thanks{}}
%

\markboth{}%
{Shell \MakeLowercase{\textit{et al.}}: Bare Demo of IEEEtran.cls for IEEE Transactions on Magnetics Journals}

\IEEEtitleabstractindextext{%
\begin{abstract}
\emph{Abstract}\text{---}Despite the increasing popularity and successful examples of crowdsourcing, it is stripped of aureole when collective efforts are derailed or severely hindered by elaborate sabotage. A service exchange dilemma arises when non-cooperation among self-interested users, and zero social welfare is obtained at myopic equilibrium. Traditional rating protocols are not effective to overcome the inefficiency of the socially undesirable equilibrium due to specific features of crowdsourcing: a large number of anonymous users having asymmetric service requirements, different service capabilities, and dynamically joining/leaving a crowdsourcing platform with imperfect monitoring. In this paper, we develop the first game-theoretic design of the two-sided rating protocol to stimulate cooperation among self-interested users, which consists of a recommended strategy and a rating update rule. The recommended strategy recommends a desirable behavior from three predefined plans according to intrinsic parameters, while the rating update rule involves the update of ratings of both users, and uses differential punishments that punish users with different ratings differently. By quantifying necessary and sufficient conditions for a sustainable social norm, we formulate the problem of designing an optimal two-sided rating protocol that maximizes the social welfare among all sustainable protocols, provide design guidelines for optimal two-sided rating protocols and a low-complexity algorithm to select optimal design parameters in an alternate manner. Finally, evaluation results show the validity and effectiveness of our protocol designed for service exchange dilemma in crowdsourcing.
\end{abstract}

\begin{IEEEkeywords}
service exchange, incentive mechanism, rating protocol, differential punishment, sustainable social norm, game theory
\end{IEEEkeywords}}

\maketitle

\IEEEdisplaynontitleabstractindextext

%
\IEEEpeerreviewmaketitle

\section{INTRODUCTION}
%
%
%
%

\IEEEPARstart{C}{rowdsourcing} has emerged in recent years as a paradigm for leveraging human intelligence and activity at a large scale, it offers a distributed and cost-effective approach to obtain needed content, information or services by soliciting contributions from an undefined set of people, instead of assigning a job to designated employees \cite{1,101}. Over the past decade, numerous successful crowdsourcing platforms, such as Amazon Mechanical Turk (AMT) \cite{102}, Yahoo! Answers \cite{R1}, Upwork \cite{RR1} emerge. With the help of crowdsourcing platforms and with the power of a crowd, crowdsourcing is becoming increasingly popular as it provides an efficient and cheap method of obtaining solutions to complex tasks that are currently beyond computational capabilities but possible for humans \cite{R2,R3,R4}.

Over the past decade, techniques for securing crowdsourcing operations have been expanding steadily, so has the number of applications of crowdsourcing \cite{2}. However, users in a crowdsourcing platform have the opportunity to exhibit antisocial behaviors due to the openness of crowdsourcing\textcolor{black}{, and hence crowdsourcing is stripped of aureole when collective efforts are derailed or severely hindered by elaborate sabotage \cite{Naroditskiy,3}.} As part of crowdsourcing, service exchange applications have proliferated as the medium that allows users to exchange valuable services. In a typical service exchange application, a user plays a dual role: \textcolor{black}{as a client who submits his requirement to a crowdsourcing platform, and as a server who chooses to devote a high/low level of efforts to work on a job and provides solutions to the client in exchange for rewards \cite{27}.} Since providing services incurs costs to servers in terms of power, time, bandwidth, privacy leakage, etc., rational and self-interested users would be more inclined to devote low level efforts when being a server, and seek for services from others as a client, rather than providing services as a server. Under such circumstances, non-cooperative behaviors among self-interested users decrease their social welfare, which is a social dilemma. Therefore, an increased level of cooperation is considered to be socially desirable for service exchange in crowdsourcing platforms.

The main reason why users in the above service exchange game have the incentive not to cooperate with each other is the absence of punishments for such malicious behaviors. Self-interested users always adjust their strategies over time to maximize their own utilities, however, they cannot receive a direct and immediate benefit by choosing to be a server and devoting a high-level effort to provide high-quality services to other users (as clients). Such a conflict leads to an inevitable fact that, many users would be apt to be a client to request services, or be apt to be a server but devote a low-level effort to provide low-quality services. Thus, an important functionality of the crowdsourcing platform is to provide a good incentive mechanism for service exchange. And there is an urgent need to stimulate cooperation among self-interested users in crowdsourcing, under which self-interested users will be compelled to follow the social norm such that the inefficiency of the socially undesirable equilibrium will be overcome, \emph{i.e.}, if a user chooses to be a server in the first stage, and provides high-quality services in the second stage, then he should be rewarded immediately, otherwise, he should be punished \cite{6}.

Incentives are key to the success of crowdsourcing as it heavily depends on the level of cooperation among self-interested users. There are two types of incentives, \emph{monetary} and \emph{non-monetary}. Incentive mechanisms based on monetary incentivize individuals to provide high-quality services relying on monetary or matching rewards in the form of micropayments, which in principle can achieve the social optimum by internalizing external effects of self-interested individuals. \textcolor{black}{The work \cite{icis09} presents a game theoretic model of an all-pay contest in crowdsourcing, and investigates whether multiple prizes can maximize contest revenue. }Although monetary incentives, in some sense, are the best and easiest way to motivate people \cite{anand}, several challenges prevent monetary incentives from success in service exchange applications. Firstly, it is difficult to price small services (\emph{e.g.}, answer, knowledge, resources etc.) being exchanged between users as these are not real goods \cite{23}. Deploying auctions to set the price may reduce the price to a certain degree, while it may cause implementation complexity, high delay, and currency inflation \cite{dd}. Secondly, as pointed out by \cite{8A}, \cite{12} and \cite{12B}, ``free-riding'' may happen when rewards are paid before providing services, a server always has the incentive to take the reward without devoting enough effort, whereas if rewards are paid after the service exchange is completed, ``false-reporting'' may arise since the client has an incentive to lower or refuse rewards to servers by lying about the outcome of the task. Thirdly, although a monetary scheme is simple to be designed, it often requires a complex accounting infrastructure, which introduces computation overheads and substantial communication, and thus difficult to be implemented in reality \cite{luo,singh}.

In addition to monetary incentives, some applications are endowed with different non-monetary incentive types, such as natural incentives, personal development, solidary incentives, material incentives, etc. \cite{anand}. Among these non-monetary incentives, rating protocols (as a form of solidary incentives) originally proposed by Kandori \cite{24} have been shown to work effectively as incentive mechanisms to force cooperation in crowdsourcing platforms \cite{6,23,26,27,xie}. Generally speaking, a rating protocol labels each user by a rating label based on his past behaviors indicating his social status in the system. And users with different ratings are treated differently by the other users they interact with, and the rating of a user who complies with (\emph{resp.} deviates from) the social norm goes up (\emph{resp.} down). Hence, a user with high/low rating can be rewarded/punished by other users in a crowdsourcing platform who have not had past interactions with him. Furthermore, the use of ratings as a summary record of a user requires significantly less amount of information being maintained \cite{adamic}. Hence, the rating protocol has a potential to form a basis for successful incentive mechanisms for service exchange in crowdsourcing platforms. Motivated by the above considerations, this paper is devoted to the study of incentive mechanisms based on rating protocol.

However, there are several major reasons that prevent existing works on the rating protocol to be directly implemented for incentive provision for service exchange in crowdsourcing: (\emph{\romannumeral 1}) Users have asymmetric service requirements and they can freely and frequently change their partners they interact with in most crowdsourcing platforms, which results in asymmetric interactions among those users, and it is more difficult to model and analyze \cite{hew,gupta};
(\emph{\romannumeral 2}) Taking into account the service capability of users and the spatial/temporal requirements of tasks, using the framework of anonymous random matching games in which each user is repeatedly matched with different partners over time for service exchange is inappropriate \cite{27,23};
(\emph{\romannumeral 3}) User population is large, users are anonymous and not sufficiently patient, especially when those users with bad ratings may attempt to leave and rejoin the system as new members to avoid punishments (\emph{i.e.}, whitewashing)\cite{xie,feldman};
(\emph{\romannumeral 4}) In the presence of imperfect monitoring, a user's rating may be wrongly updated, which will impact on rating protocol design, as well as social welfare loss \cite{6,26}.

In this paper, we take into account the above features of service exchange in crowdsourcing into consideration, and propose a game-theoretic framework for designing and analyzing a class of rating protocols based incentive mechanisms, in order to stimulate cooperation among self-interested users and maximize the social welfare. \textcolor{black}{To the best of our knowledge, the update of ratings of both users (we name it as a two-sided rating) matched in the service exchange game is rarely tackled in the literature.} Using game theory to analyze how cooperation can be enforced and how to maximize the social welfare under the designed two-sided rating protocol, we rigorously analyze how users' behaviors are influenced by intrinsic parameters and design parameters as well as users' evaluation of their individual long-term utilities, in order to characterize the optimal design that maximizes users' utilities and enforces cooperation among them. The main contributions of this paper are summarized as follows:

\begin{figure*}
\centering
\includegraphics[height=5cm ,width=15cm,angle=0,scale=1]{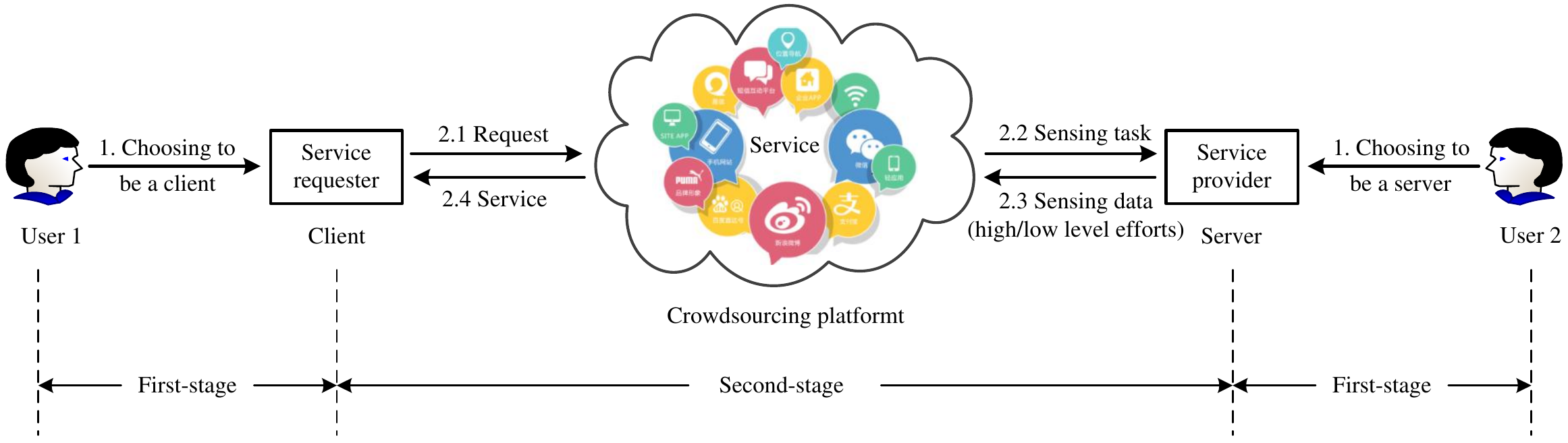}
\caption{\textcolor{black}{A general crowdsourcing based service exchange model}}
\label{model}
\end{figure*}

\begin{enumerate}[(i)]
 \item  We model the service exchange problem as an asymmetric game model with two stages, and show that inefficient outcome arises when no user cooperates with each other, and thus zero social welfare is obtained at myopic equilibrium, which is a social dilemma.
 \item  We develop the first game-theoretic design of two-sided rating protocols to stimulate cooperation among self-interested users, which consists of a recommended strategy and a rating update rule. The recommended strategy recommends a desirable behavior chosen from three predefined recommended plans according to intrinsic parameters, while the rating update rule involves the update of ratings of both users, and uses differential punishments that punish users with different ratings differently.
 \item We formulate the problem of designing an optimal two-sided rating protocol that maximizes the social welfare among all sustainable rating protocols, provide design guidelines for determining whether there exists a sustainable two-sided rating protocol under a given recommended strategy, and design an algorithm achieving low-complexity computation via a two-stage procedure, each stage consists of two steps (we call this a two-stage two-step procedure), in an alternate manner.
 \item We use simulation results to demonstrate how intrinsic parameters (\emph{i.e.}, costs, imperfect monitoring, user's patience) impact on optimal recommended strategies, the design parameters to characterize the optimal design of various protocols, and the performance gain of the proposed optimal two-sided rating protocol.
\end{enumerate}

The remainder of this article is organized as follows. In section II, we describe the service exchange dilemma game with two-sided rating protocols. In section III, we formulate the problem of designing an optimal two-sided rating protocol. Then we provide the optimal design of two-sided rating protocols in Section IV. \textcolor{black}{Section V presents evaluation results to illustrate key features of the designed protocol.} Finally, conclusions are drawn in Section VI.

\section{SYSTEM MODELS}

\subsection{Service Exchange Dilemma Game}

\textcolor{black}{As illustrated in Figure \ref{model}, a crowdsourcing based service exchange system consisting of a platform with several users on the Internet, where each user in a crowdsourcing platform can offer services to other users. Examples of services include sensing tasks, expert knowledge, information resource, computing power, storage space, etc. The crowdsourcing process can be described as follow: On the one hand, each user can choose to become either a service requester (\emph{i.e.}, client) or a service provider (\emph{i.e.}, server). On the other hand, a client generates a service request which is sent to a matched server, and the server devotes a high/low level of efforts to provide the requested service to the client.}

We model such a process using uniform random matching, that is each user in the community is involved in two matches in every period, one as a client and the other as a server, each user is equally likely to receive exactly one request in every period, and the matching is independent in different periods. Note that the user with whom a user interacts as a client can be different from that with whom he interacts as a server, reflecting asymmetric interests between a pair of users in a given instant. Such a model well approximates the matching process between users in large-scale crowdsourcing systems where users interact with others in an ad-hoc fashion and the interactions between users are constructed randomly over time.

In this model, a user decides whether or not to request service (\emph{choosing to be a client/server}), if the user chooses to be a server, he strategically determines his service quality (\emph{devoting a high/low level of efforts}). Note that the decisions are sequential: the decision on role selection is made first, and then the decision on service quality is made next. We model this interaction as a sequential game. Formally, we have a two-stage game. In the first stage, a user's action is chosen from the set $\{C,S\}$, where $C$ stands for ``choosing to be a client'' (request service), whereas $S$ stands for ``choosing to be a server''(offer service). In the second stage, the server has a binary choice of whether being whole-hearted or being half-hearted in providing the service, while the client has no choice. The set of actions for the server is denoted by $\{H,L\}$ , where $H$ stands for ``high level of effort'' and $L$ stands for ``low level of effort''.

We assume that any $C$ strategy is costly (consumes a cost $c$ for choosing $C$). If the server devotes a high level of efforts to fulfill the client's request, the client receives a service benefit of $b > 0$, while the server suffers a service cost of $s > 0$. If the server devotes a low level of efforts to the request, both users receive zero payoffs. Obviously, the server's action determines the payoffs of both users.  After a server takes an action, the client sends a report about the action of the server to the third-party device or infrastructure that manages the rating scores of users. \textcolor{black}{Taking into account imperfect monitoring, the report is inaccurate, either by the client's incapability of accurate assessment or by some system error with a small probability $\varepsilon$.} That is, $L$ is reported when the server takes action $H$ with probability $\varepsilon$, and vice versa\footnotemark[1]\footnotetext[1]{\textcolor{black}{In this paper, we focus on the situation in which probability that errors occur in the first-stage of the game is approximated by 0, because when the probability for erroneous report of $C$ or $S$ is very small, errors occurring in the first stage is easy to be detected and corrected in time.}}. Assuming a binary set of reports, it is without loss of generality to restrict $\varepsilon\in[0, \frac{1}{2})$, because when $\varepsilon=\frac{1}{2}$, reports are completely random and do not contain any meaningful information about the actions of users. \textcolor{black}{Conveniently, Table \ref{table00} lists frequently used notions in this paper.}

We find the subgame perfect equilibrium of the two-stage game. Each pair of users' decisions made in the first stage ($C$ or $S$) result in a different second-stage game ($H$ or $L$). We first compute expected utilities in the second-stage game, and then turn back to compute expected utilities when both users choose their actions in the first-stage before knowing their productivities. \textcolor{black}{The pay-off matrix for the game played in the first stage was depicted in Table \ref{tab02}. The detailed computation process is in Appendix A.}

In summary, for any choice of parameters, only $SS$ can be a Nash equilibrium of the service exchange game. When every user chooses his action to maximize his current payoff myopically, an inefficient outcome arises where every user receives zero payoff, which is a social dilemma. Under the current framework, nobody will take the initiative to help others, and do not expect to get help from others.

\begin{table}[tb]
\scriptsize
\caption{\textcolor{black}{SUMMARY OF NOTATIONS IN THIS PAPER}}
\label{table00}
\begin{center}
\begin{tabular}{|c|c|c|c|c|c|c|}
\hline

  \multicolumn{1}{c|}{\textbf{\textcolor{black}{Notations}}}& \multicolumn{6}{|c}{\textbf{\textcolor{black}{Physical Meanings}}} \\
\hline
\multicolumn{1}{c|}{\textcolor{black}{$c$}}& \multicolumn{6}{|l}{\textcolor{black}{cost for choosing to be a client.}}\\
\hline
\multicolumn{1}{c|}{\textcolor{black}{$s$}}& \multicolumn{6}{|l}{\textcolor{black}{cost for devoting a high level of effort to fulfill the client's request.}}\\
\hline
\multicolumn{1}{c|}{\textcolor{black}{$b$}}& \multicolumn{6}{|l}{\textcolor{black}{service benefit if the service request be fulfilled.}}\\
\hline
\multicolumn{1}{c|}{\textcolor{black}{$\varepsilon$}}& \multicolumn{6}{|l}{\textcolor{black}{probability that errors occur in the second-stage game.}}\\
\hline
\multicolumn{1}{c|}{\textcolor{black}{$\omega$}}& \multicolumn{6}{|l}{\textcolor{black}{discount factor to denote users' patience.}}\\
\hline
\multicolumn{1}{c|}{\textcolor{black}{$\mathcal P$}}& \multicolumn{6}{|l}{\textcolor{black}{rating protocol.}}\\
\hline
\multicolumn{1}{c|}{\textcolor{black}{$\theta$}}& \multicolumn{6}{|l}{\textcolor{black}{rating label.}}\\
\hline
\multicolumn{1}{c|}{\textcolor{black}{$\Theta$}}& \multicolumn{6}{|l}{\textcolor{black}{set of rating labels.}}\\
\hline
\multicolumn{1}{c|}{\textcolor{black}{$\sigma$}}& \multicolumn{6}{|l}{\textcolor{black}{social strategy.}}\\
\hline
\multicolumn{1}{c|}{\textcolor{black}{$\pi$}}& \multicolumn{6}{|l}{\textcolor{black}{recommended strategy for a server.}}\\
\hline
\multicolumn{1}{c|}{\textcolor{black}{$\tau$}}& \multicolumn{6}{|l}{\textcolor{black}{rating update rule.}}\\
\hline
\multicolumn{1}{c|}{\textcolor{black}{$\alpha_{\theta}$}}& \multicolumn{6}{|l}{\textcolor{black}{strength of reward imposed to a server with rating $\theta$.}}\\
\hline
\multicolumn{1}{c|}{\textcolor{black}{$\beta_{\theta}$}}& \multicolumn{6}{|l}{\textcolor{black}{strength of punishment imposed to a server with rating $\theta$.}}\\
\hline
\multicolumn{1}{c|}{\textcolor{black}{$\gamma_{\theta}$}}& \multicolumn{6}{|l}{\textcolor{black}{strength of reward imposed to a client with rating $\theta$.}}\\
\hline
\multicolumn{1}{c|}{\textcolor{black}{$\delta_{\theta}$}}& \multicolumn{6}{|l}{\textcolor{black}{strength of punishment imposed to a client with rating $\theta$.}}\\
\hline
\multicolumn{1}{c|}{\textcolor{black}{$\rho$}}& \multicolumn{6}{|l}{\textcolor{black}{ratio of the number of a user choosing to be a client and a server.}}\\
\hline
\multicolumn{1}{c|}{\textcolor{black}{$r$}}& \multicolumn{6}{|l}{\textcolor{black}{reported service quality by a client.}}\\
\hline
\multicolumn{1}{c|}{\textcolor{black}{$q$}}& \multicolumn{6}{|l}{\textcolor{black}{actual service quality devoted by a server.}}\\
\hline
\multicolumn{1}{c|}{\textcolor{black}{$\eta_\theta$}}& \multicolumn{6}{|l}{\textcolor{black}{stationary distribution of rating labels.}}\\
\hline
\multicolumn{1}{c|}{\textcolor{black}{$v$}}& \multicolumn{6}{|l}{\textcolor{black}{expected one-period utility of a user.}}\\
\hline
\multicolumn{1}{c|}{\textcolor{black}{$v^\infty$}}& \multicolumn{6}{|l}{\textcolor{black}{expected long-term utility of a user.}}\\
\hline
\multicolumn{1}{c|}{\textcolor{black}{$U_{\mathcal P}$}}& \multicolumn{6}{|l}{\textcolor{black}{social utility under the rating protocol $\mathcal P$}}\\
\hline
\end{tabular}%
\end{center}
\end{table}

\begin{table}[tb]

\caption{The expected payoff matrix for the first stage game}
\label{tab02}
\begin{center}
\begin{tabular}{c|c|c|c|c|c|c}

\hline
  \multicolumn{1}{c|}{\textbf{}}& \multicolumn{3}{c|}{\textbf{$C$}}& \multicolumn{3}{c}{\textbf{$S$}} \\
\hline
  \multicolumn{1}{c|}{\textbf{$C$}}& \multicolumn{3}{c|}{\quad\quad\textbf{$-c$, $-c$}\quad\quad\quad}& \multicolumn{3}{c}{\quad\quad\textbf{$\varepsilon b-c$, $0$}\quad\quad\quad} \\
\hline
  \multicolumn{1}{c|}{\textbf{$S$}}& \multicolumn{3}{c|}{\quad\quad\textbf{$0$, $\varepsilon b-c$}\quad\quad\quad}& \multicolumn{3}{c}{\quad\quad\textbf{$0$, $0$}\quad\quad\quad} \\
\hline

\end{tabular}%
\end{center}
\end{table}
\subsection{Two-sided Rating Protocols}

We consider a two-sided rating protocol that consists of a recommended strategy and a rating update rule. The recommended strategy prescribes the contingent plan according to intrinsic parameters that the server should take based on ratings of both his own and his client's. Here, we focus on one plan, while two other plans will be introduced in the later half of this article. The rating update rule involves the update of ratings of both users depending on their past actions as a server or a client, and uses differential punishments that punish users with different ratings differently. To the best of our knowledge, two-sided rating protocol in crowdsourcing is rarely tackled in the literature. In the following, we give a formal definition of a two-sided rating protocol.

\begin{myDef}
A two-sided rating protocol $\mathcal{P}$ is represented as a 5-tuple $(\Theta,\sigma,\rho,\pi,\tau)$, i.e., a set of binary rating labels $\Theta$, a social strategy $\sigma$, a client/server ratio $\rho$, a recommended strategy $\pi$, and a rating update rule $\tau$.
\begin{itemize}
       \item 	$\Theta=\{0,1\}$ denotes the set of binary rating labels, where 0 is the bad rating, and 1 is the good rating.
       \item    $\sigma:\Theta \rightarrow \mathcal{A}$ represents the adopted social strategy for a user with rating $\theta$, where $\sigma(\theta|\theta \in \Theta )\in \{\{C,S\}\times \{H,L\}\}$.
       \item    \textcolor{black}{$\rho:\Theta \rightarrow R^+$ denotes the ratio that a user with rating $\theta$ chooses to become a client and a server, which contains his historical choice and current choice of $\sigma(\theta|\theta \in \Theta )\in \{C,S\}$.}
       \item 	$\pi:\Theta\times\Theta \rightarrow \mathcal{A} $ defines the strategy $\sigma(\theta_S,\theta_C)\in \mathcal{A}$ which the server with rating $\theta_S$ should select when faced with the client with rating $\theta_C$.
 \begin{eqnarray}
 {\pi(\theta_S,\theta_C)=}
  \left \{
  \begin{array}{ll}
      1,  & if~ \theta_S \leq \theta_C\\
      0,  & otherwise
  \end{array}
  \right.
\end{eqnarray}
       \item \textcolor{black}{$\tau$ can be denoted by a tuple $(\tau_S,\tau_C)$, where $\tau_S:\Theta \times \mathcal{A}\rightarrow \Delta(\Theta)$ updates the rating of a server based on his current rating, his matched client's rating, the reported strategy and the recommended strategy as follows:	}
\begin{eqnarray}{\textcolor{black}{\tau_S (\theta'_S| \theta_S, \theta_C, r, \pi)=}}
\label{updates}
      \left \{
\begin{array}{ll}
 \textcolor{black}{\alpha_{\theta_S},}  &\textcolor{black}{\theta'_S=1,  r\geq}\\&\textcolor{black}{\pi(\theta_S, \theta_C)} \\
  \textcolor{black}{1-\alpha_{\theta_S},}  &\textcolor{black}{\theta'_S=0, r\geq}\\&\textcolor{black}{\pi(\theta_S, \theta_C)} \\
  \textcolor{black}{\beta_{\theta_S},}  &\textcolor{black}{\theta'_S=0, r<}\\&\textcolor{black}{\pi(\theta_S, \theta_C)} \\
  \textcolor{black}{1-\beta_{\theta_S},}  &\textcolor{black}{\theta'_S=1, r<}\\&\textcolor{black}{\pi(\theta_S, \theta_C)}
\end{array}
\right.
\end{eqnarray}

\textcolor{black}{$\tau_C:\Theta \times \mathcal{A}\rightarrow \Delta(\Theta)$ specifies how a client's rating should be updated based only on his current rating as follows:}
\begin{eqnarray}{\textcolor{black}{\tau_C (\theta'_C|\theta_C, \rho)=}}
\label{updatec}
      \left \{
\begin{array}{ll}
   \textcolor{black}{\gamma_{\theta_C},}  &\textcolor{black}{\theta'_C=1,} \\&\textcolor{black}{\rho(\theta_C)\leq 1}\\
  \textcolor{black}{1-\gamma_{\theta_C},}  &\textcolor{black}{\theta'_C=0, }\\&\textcolor{black}{\rho(\theta_C)\leq 1}\\
  \textcolor{black}{\delta_{\theta_C},}  &\textcolor{black}{\theta'_C=0,} \\&\textcolor{black}{\rho(\theta_C)> 1}\\
  \textcolor{black}{1-\delta_{\theta_C},} &\textcolor{black}{\theta'_C=1,} \\&\textcolor{black}{\rho(\theta_C)> 1}
\end{array}
\right.
\end{eqnarray}

We characterize the erroneous report by a mapping $R:\{0, 1\} \rightarrow \Delta(\{0, 1\})$, where 0 and 1 represent ``L'' and ``H'', respectively. $\Delta(\{0, 1\})$ is the probability distribution over $\{0, 1\}$, and $R(r|q)$ is the probability that the client reports received service quality ``r'' given the server's provided service quality ``q''.
\begin{eqnarray} {R(r|q)=}
        \left \{
\begin{array}{ll}
  1-\varepsilon,  & r=q\\
  \varepsilon,  & r\ne q
\end{array}
\right.
\forall r,q \in\{0,1\}.
\end{eqnarray}
\end{itemize}
\end{myDef}

\begin{figure}
\centering
\includegraphics[height=3cm ,width=9cm,angle=0,scale=1]{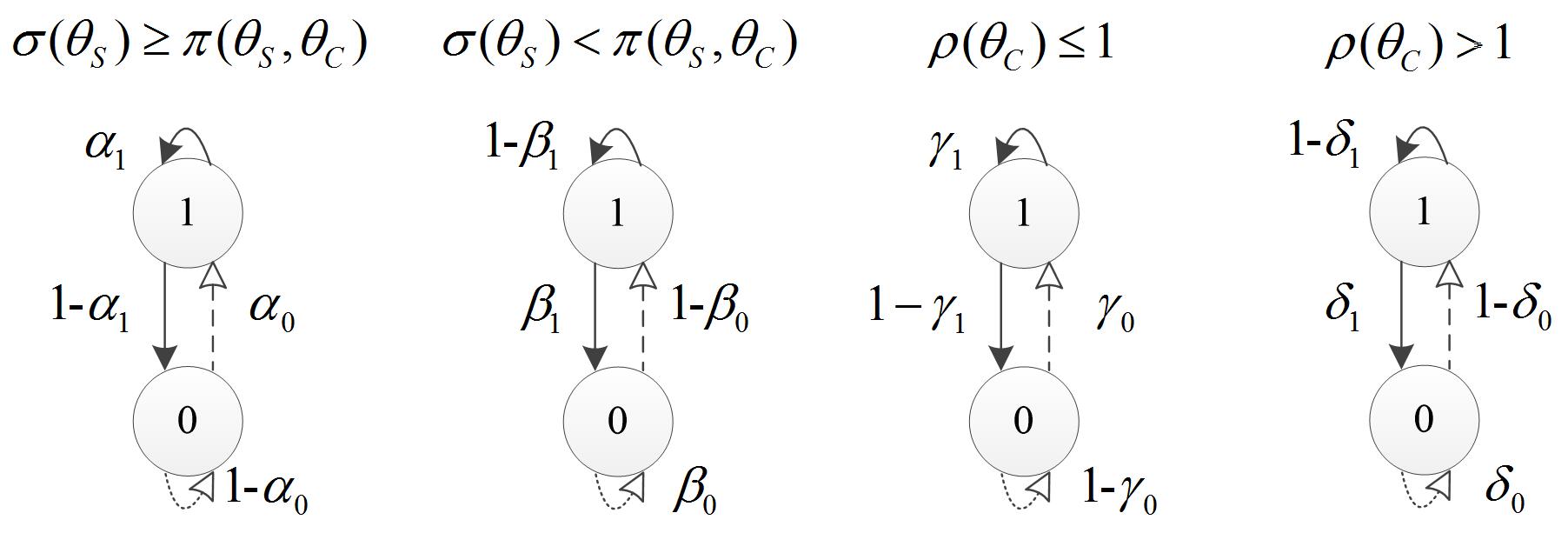}
\caption{Schematic representation of a rating update rule $\tau$}
\label{fig01}
\end{figure}

\emph{\textbf{\textcolor{black}{Remark}}}: A schematic representation of a rating update rule $\tau$ is provided in Figure \ref{fig01}. Given a rating protocol $\mathcal{P}$, each user $i$ is tagged with a binary rating label $\theta_i \in \Theta\triangleq\{0,1\}$ representing its social status. Obviously, the higher $\theta_i$ is, the better the social status the user $i$ has. Ratings of users are stored and updated by the system administrator based on strategies adopted by the user in the transactions that he is engaged in. The rating scheme $\tau$ can update a user's rating at the end of each transaction or at the beginning of the next transaction. Under the rating update rule (\ref{updates}) and (\ref{updatec}), a $\theta_S$-server (\emph{i.e.}, \textcolor{black}{a server with rating $\theta_S$) will have rating 1 with probability $\alpha_{\theta_S}$, and have rating 0 with probability $1-\alpha_{\theta_S}$, if the service quality $r$ reported by the client is no lower than the recommended service quality $\pi(\theta_S, \theta_C)$;} otherwise, it will have rating 1 with probability $\beta_{\theta_S}$, and have rating 0 with probability $1-\beta_{\theta_S}$. \textcolor{black}{Similarly, a $\theta_C$-client will have rating 1 with probability $\gamma_{\theta_C}$ and have rating 0 with probability $1-\gamma_{\theta_C}$ if the ratio $\rho\leq 1$;} otherwise, it will have rating 1 with probability $\delta_{\theta_C}$, and have rating 0 with probability $1-\delta_{\theta_C}$. Obviously,  $\alpha_{\theta_S}$ and $\gamma_{\theta_C}$ can be referred to as the strength of reward imposed on servers and clients when they cooperate with each other, respectively, while  $\beta_{\theta_S}$ can be referred to as the strength of punishment imposed on servers when they do not offer high level efforts, similarly, $\delta_{\theta_C}$ can be referred to as the strength of punishment imposed on clients when they expect to get excessive service from others rather than to serve others.

\textcolor{black}{Definition 1 describes a simple two-sided rating protocol which assigns binary rating labels to users, and provides a binary choice of whether devoting a high level of effort or a low level of effort in providing the service. Although other more elaborate two-sided rating protocols (as discussed in Section \uppercase\expandafter{\romannumeral6}) may be considered, we show that this simple one is effective to overcome the inefficiency of the service exchange dilemma in crowdsourcing.}

\section{PROBLEM FORMULATION}

\subsection{Stationary Rating Distribution}
Given a two-sided rating protocol $\mathcal{P}$, suppose that each user always follows a given recommended strategy $\pi$ and keep $\rho \leq 1$ in any period. As time passes, ratings of users are updated, and thus the distribution of users' ratings in a system evolves over time. \textcolor{black}{Let $\eta_\mathcal{P}^t(\theta)$ be the fraction of $\theta$-users in the total population at the beginning of time period \emph{t}, }then the transition from $\eta_\mathcal{P}^t(\theta)$ to $\eta_\mathcal{P}^{t+1}(\theta)$ is determined by the rating update rule $\tau$, taking into account the rate $\lambda$ for a user choosing to be a client and the error probability $\varepsilon$, as shown in the following expressions:
     \begin{equation}
  \left\{
  \begin{aligned}
    \eta_\mathcal{P}^{t+1}(0)= & \lambda[(1-\gamma_1)\eta_\mathcal{P}^{t}(1)+(1-\gamma_0)\eta_\mathcal{P}^{t}(0))]+(1-\lambda)\\
                     &\big\{(1-\varepsilon)\lbrack(1-\alpha_1)\eta_\mathcal{P}^{t}(1)+(1-\alpha_0)\eta_\mathcal{P}^{t}(0)\rbrack+\\
                     &\varepsilon\lbrack\beta_1\eta_\mathcal{P}^{t}(1)+\beta_0\eta_\mathcal{P}^{t}(0)\rbrack\big\}\\
     \eta_\mathcal{P}^{t+1}(1)= & \lambda\lbrack\gamma_1\eta_\mathcal{P}^{t}(1)+\gamma_0\eta_\mathcal{P}^{t}(0)\rbrack+(1-\lambda)\\
                     & \big\{(1-\varepsilon)\lbrack\alpha_1\eta_\mathcal{P}^{t}(1)+\alpha_0\eta_\mathcal{P}^{t}(0)\rbrack+\\
                     & \varepsilon\lbrack(1-\beta_1)\eta_\mathcal{P}^{t}(1)+(1-\beta_0)\eta_\mathcal{P}^{t}(0)\rbrack\big\}
 \end{aligned}
  \right.
  \end{equation}

Here we set $\phi_1=\lambda(1-\gamma_1)+(1-\lambda)[(1-\varepsilon)(1-\alpha_1)+\varepsilon\beta_1]$, $\phi_0=\lambda(1-\gamma_0)+(1-\lambda)[(1-\varepsilon)(1-\alpha_0)+\varepsilon\beta_0]$, the stationary distribution $\{\eta_\mathcal{P} (\theta)\}_{\theta=0}^1$ can be derived as follows.
  \begin{equation} {}
  \label{a1}
  \left\{
  \begin{aligned}
     & \eta_\mathcal{P} (0)=\frac{\phi_1}{1+\phi_1-\phi_0}\\
     & \eta_\mathcal{P} (1)=\frac{1-\phi_0}{1+\phi_1-\phi_0}\\
 \end{aligned}
  \right.
  \end{equation}

Since the coefficients in the equations that define a stationary distribution are independent of the recommended strategy that users should follow, the stationary distribution is also independent of the recommended strategy, as can be seen from Eq.(\ref{a1}). Thus, we will write the stationary distribution as $\{\eta_\mathcal{P} (\theta)\}$ to emphasize its dependence on $\mathcal{P}$.

\subsection{Sustainable Conditions}

The purpose of designing a social norm is to enforce a user to follow the recommended strategy $\pi(\theta_S, \theta_C)$ and keep $\rho \leq 1$ in any period. We call a user who complies with such a social norm as a ``compliant user'', otherwise, the user who deviates from the social norm is called as a ``non-compliant user''. The compliant user will be rewarded, on the contrary, a non-compliant user will be punished in order to regulate his behavior. Since we consider a non-cooperative scenario, it is important to check whether a user can improve his long-term payoff by a unilateral deviation. Note that any unilateral deviation from an individual user would not affect the evolution of rating scores and thus the stationary distribution, because we consider a continuum of users.

\textcolor{black}{Let $c_{\pi}(\widetilde{\theta}, \theta)$ be the cost paid by a $\widetilde{\theta}$-server who is matched with a $\theta$-client and follows a recommended strategy $\pi$, that is, $c_{\pi}(\widetilde{\theta}, \theta) = s$ if  $\pi(\widetilde{\theta}, \theta)= 1$, and $c_{\pi}(\widetilde{\theta}, \theta) = 0$ if  $\pi(\widetilde{\theta}, \theta)= 0$. Similarly, let $b_{\pi}(\widetilde{\theta}, \theta)$ be the benefit received by a $\theta$-client who is matched with a $\widetilde{\theta}$-server following a recommended strategy  $\pi$, that is, $b_{\pi}(\widetilde{\theta}, \theta) = b-c$ if  $\pi(\widetilde{\theta}, \theta) = 1$ and $b_{\pi}(\widetilde{\theta}, \theta) =-c$ if  $\pi(\widetilde{\theta}, \theta) = 0$. Since we consider uniform random matching, the expected one period payoff of a $\theta$-user under a rating protocol $\mathcal{P}$ and a chosen rate $\lambda$ before he is matched is given by}
\begin{equation}
\label{pp}
\begin{aligned}
v_{\mathcal{P},\lambda}(\theta)=\lambda\sum_{\widetilde{\theta}\in\Theta}{\eta(\widetilde{\theta})b_{\pi}(\widetilde{\theta}, \theta)}-(1-\lambda)\sum_{\widetilde{\theta}\in\Theta}{\eta(\widetilde{\theta})c_{\pi}(\theta,\widetilde{\theta})}\\
\end{aligned}
\end{equation}

To evaluate the long-term payoff of a compliant user, we use the discounted sum criterion in which the long-term payoff of a user is given by the expected value of the sum of discounted period payoffs starting from the current period. Let $p_\mathcal{P}(\theta'|\theta)$ be the transition probability that a $\theta$-user becomes a $\theta'$-user in the next period under a rating protocol $\mathcal{P}$  when he follows the recommended strategy and selects the chosen rate $\lambda$, which can be expressed as
\begin{eqnarray}  {p_{\mathcal{P},\lambda}(\theta'|\theta)=}
\label{two}
  \left \{
\begin{array}{ll}
      \lambda\gamma_{\theta}+(1-\lambda) [(1-\varepsilon)\alpha_{\theta}+\\ \varepsilon(1-\beta_{\theta})], &if \theta'=1\\
      \lambda(1-\gamma_{\theta})+(1-\lambda)\\ \lbrack(1-\varepsilon)(1-\alpha_{\theta})+\varepsilon\beta_{\theta}\rbrack, &if \theta'=0\\
\end{array}
\right.
\end{eqnarray}

\textcolor{black}{The expected long-term utility of a $\theta$-user is the infinite horizon discounted sum of his expected one-period utility with his expected future payoff multiplied by a common discount factor $\omega$, which can be computed by solving the following recursive equation:}
\begin{equation} {}
\label{lpv}
v_{\mathcal{P},\lambda}^\infty (\theta)= v_{\mathcal{P},\lambda} (\theta)+ \omega \sum_{\theta' \in \Theta} p_{\mathcal{P},\lambda} (\theta' |\theta)v_{\mathcal{P},\lambda}^\infty (\theta')
\end{equation}
Where $\omega\in[0,1)$ is the rate at which a user discounts his future payoff, and reflects his patience\footnotemark[2]\footnotetext[2]{\textcolor{black}{It is obvious that a larger discount factor reflects a more patient user, but no user is patient with a discount factor $\omega=1$ as no one is willing to stay in a system forever.}}.

\textcolor{black}{Together with Eq.(\ref{two}) and Eq.(\ref{lpv}), we have Eq.(\ref{bp20}), the detailed computation process is in Appendix B.}
\begin{equation} {}
  \label{bp20}
  \left\{
  \begin{aligned}
  \begin{split}
  &v_\mathcal{P,\lambda}(0)=\lambda\big\{[\eta_\mathcal{P}(0)(1-\varepsilon)+\eta_\mathcal{P}(1)\varepsilon]b-c\big\}-(1-\lambda)s\\
  &v_\mathcal{P,\lambda}(1)=\lambda[(1-\varepsilon)b-c]-(1-\lambda)\eta_\mathcal{P}(1)s\\
  &v_\mathcal{P}^\infty (1)-v_\mathcal{P}^\infty (0)=\frac{\lambda\eta_\mathcal{P}(1)(1-2\varepsilon)b+(1-\lambda)\eta_\mathcal{P}(0)s}{1+\omega(\phi_1-\phi_0)}
 \end{split}
 \end{aligned}
 \right.
 \end{equation}

\textcolor{black}{It is surprising that $v_\mathcal{P}^\infty (1)-v_\mathcal{P}^\infty (0)$ is a constant given $\eta_\mathcal{P}(\theta)$, which is very convenient for optimal designing the proposed two-sided rating protocols in the remainder of this paper.} Since users always aim to strategically maximize their own benefits, they will find in their own self-interest to comply with the social norm under a given two-sided rating protocol, if and only if they cannot benefit in terms of their long-term utilities upon deviations. We call such a protocol as a sustainable two-sided rating protocol, and give its formal definition as follows:

\begin{myDef}
(\textbf{Sustainable Two-sided Rating Protocols}) \textcolor{black}{A two-sided rating protocol $\mathcal{P}$ is sustainable if and only if $v_{\mathcal{P},\pi,\lambda=\frac{1}{2}}^\infty (\theta) \geq v_{\mathcal{P},\pi',\lambda'}^\infty (\theta)$, for all $\pi'\neq \pi$, $\lambda'\neq \lambda$ and $\theta\in\Theta$.}
\end{myDef}

In other words, a sustainable two-sided rating protocol $\mathcal{P}$ should maximize a user's expected long-term utility at any period, such that no user can gain from a unilateral deviation regardless of the rating of his matched partner when every other user follows the recommended strategy $\pi$ and selects $\lambda=\frac{1}{2}$. It is obvious that  the social welfare will be maximized when compliant users keep $\rho\leq 1$ (\emph{i.e.}, $\lambda\leq\frac{1}{2}$). Checking whether a rating protocol is sustainable in the second stage using the preceding definition requires computing deviation gains from all possible recommended strategies. By employing the criterion of unimprovability in Markov decision theory \cite{markov}, a user's strategic decision problem can be formulated as a Markov decision process under a two-sided rating protocol $\mathcal{P}$, where the state is the user's rating $\theta$, and the action is his chosen strategy $\sigma(\theta)$, we thus establish the one-shot deviation principle for sustainable two-sided rating protocols, \textcolor{black}{which provides simple conditions}.

\begin{myLem}
(\textbf{One-Shot Deviation Principles}) A two-sided rating protocol $\mathcal{P}$ satisfies the one-shot deviation principle if and only if
  \begin{equation} {}
  \label{osdp}
  \begin{split}
 &\frac{\frac{1}{2}[\eta_\mathcal{P}(1)(1-2\varepsilon)b+\eta_\mathcal{P}(0)s]}{1+\omega(\phi_1-\phi_0)}\geq \\
 &\max\Big\{\frac{s}{\omega(1-2\varepsilon)(\alpha_0+\beta_0-1)},\frac{\eta_{\mathcal{P}(1)}s}{\omega(1-2\varepsilon)(\alpha_1+\beta_1-1)}\Big\}
 \end{split}
  \end{equation}
\end{myLem}

\begin{IEEEproof}
For the ``if'' part: A user's expected long-term utility when he adopts the recommend strategy $\pi$ for all $\theta \in \Theta$, can be expressed as $v_{\mathcal{P}}^\infty (\theta)$ in Eq.(\ref{lpv}) (here, we fix $\lambda=\frac{1}{2}$). If the user unilaterally deviates from $\pi$ to $\pi'$ at rating $\theta$, his expected long term utility becomes
\begin{equation} {}
\label{dpv}
v_{\mathcal{P},\pi'}^\infty (\theta)= v_{\mathcal{P},\pi'} (\theta)+ \omega \sum_{\theta' \in \Theta} p_{\mathcal{P},\pi'} (\theta' |\theta,\alpha)v_{\mathcal{P},\pi'}^\infty (\theta')
\end{equation}
Where $p_{\mathcal{P},\pi'}(\theta'|\theta)$ is the transition probability that a non-compliant $\theta$-server becomes a $\theta'$-server in the next period under $\mathcal{P}$, which is expressed as
\begin{eqnarray}  {p_{\mathcal{P},\pi'}(\theta'|\theta)=}
  \left \{
\begin{array}{ll}
      \frac{1}{2}\gamma_\theta+\\ \frac{1}{2}[\varepsilon\alpha_{\theta}+(1-\varepsilon)(1-\beta_{\theta})], &if \theta'=1\\
      \frac{1}{2}(1-\gamma_\theta)+\\ \frac{1}{2}[\varepsilon(1-\alpha_{\theta})+(1-\varepsilon)\beta_{\theta}], &if \theta'=0\\
\end{array}
\right.
\end{eqnarray}
By comparing these two payoffs $v_{\mathcal{P},\pi}^\infty (\theta)$ and $v_{\mathcal{P},\pi'}^\infty (\theta)$, and solving the following inequality:
\begin{equation} {}
\begin{split}
\label{com}
&v_{\mathcal{P},\pi}^\infty (\theta)-v_{\mathcal{P},\pi'}^\infty (\theta)=\frac{1}{2}\sum_{\widetilde\theta \in \Theta}{\eta_\mathcal{P}(\widetilde\theta)(c_{\pi'}(\theta, \widetilde\theta)-c_{\pi}(\theta, \widetilde\theta))} +\\
&~~~~\omega\sum_{\theta' \in \Theta}{ [p_\mathcal{P,\pi}(\theta'|\theta)-p_{\mathcal{P},\pi'}(\theta' | \theta)]v_\mathcal{P}^\infty (\theta')}\geq 0.
\end{split}
\end{equation}
If $\theta$=0, then for each $\widetilde\theta\in\Theta$, $\pi(\theta, \widetilde\theta)=1$, $c_{\pi}(\theta, \widetilde\theta)=s$ and $c_{\pi'}(\theta, \widetilde\theta)=0$, we have
\begin{equation} {}
  \label{vp1}
  \begin{split}
  v_\mathcal{P}^\infty (1)- v_\mathcal{P}^\infty (0)\geq \frac{s}{\omega(1-2\varepsilon)(\alpha_0+\beta_0-1)}
 \end{split}
  \end{equation}
While if $\theta$=1 and $\widetilde\theta=1$, then $\pi(\theta, \widetilde\theta)=1$, $c_{\pi}(\theta, \widetilde\theta)=s$ and $c_{\pi'}(\theta, \widetilde\theta)=0$. Else if $\theta$=0, then $\pi(\theta, \widetilde\theta)=0$ for each $\widetilde\theta\in\Theta$, self-interested users have no incentive to deviate from $\pi(\theta, \widetilde\theta=0)$. Hence, we have
\begin{equation} {}
  \label{vp2}
  \begin{split}
  v_\mathcal{P}^\infty (1)- v_\mathcal{P}^\infty (0)\geq \frac{\eta_{\mathcal{P}(1)}s}{\omega(1-2\varepsilon)(\alpha_1+\beta_1-1)}
 \end{split}
  \end{equation}

We have inequality in Eq.(\ref{osdp}) by substituting $v_\mathcal{P}^\infty (1)- v_\mathcal{P}^\infty (0)=\frac{\frac{1}{2}(\eta_\mathcal{P}(1)b-\eta_\mathcal{P}(0)s)}{1+\omega(\phi_1-\phi_0)}$ into the LHS of Eq.(\ref{vp1}) and Eq.(\ref{vp2}). Hence, the two-sided rating protocol $\mathcal{P}$ is satisfied with the one-shot deviation principle if Eq.(\ref{osdp}) holds.

For the ``only if'' part: Suppose the rating protocol $\mathcal{P}$ is satisfied with the one-shot deviation principle, then clearly there are no profitable one-shot deviations. We can prove the converse by showing that if $\mathcal{P}$ is not satisfied with the one-shot deviation principle, there is at least one profitable one-shot deviation. Since $c_{\pi}(\theta, \widetilde\theta)$ and $c_{\pi'}(\theta, \widetilde\theta)$ are bounded, this is true by the unimprovability property in Markov decision theory.
\end{IEEEproof}

Lemma 1 shows that if a user cannot gain by unilaterally deviating from $\pi$ only in the current period and following $\pi$ afterwards, he can neither gain by switching to any
other recommended strategy $\pi'$, and vice versa. $\frac{1}{2}\sum_{\widetilde\theta \in \Theta}{\eta_\mathcal{P}(\widetilde\theta)(c_{\pi'}(\theta, \widetilde\theta)-c_{\pi}(\theta, \widetilde\theta))}$ of Eq.(\ref{com}) can be interpreted as the current gain from choosing $\pi'$ in the second stage, while $\omega \sum_{\theta' \in \Theta}{ [p_\mathcal{P,\pi}(\theta'|\theta)-p_{\mathcal{P},\pi'}(\theta' | \theta)]v_\mathcal{P}^\infty (\theta')}$ of Eq.(\ref{com}) represents the discounted expected future loss due to the different transition probabilities incurred by choosing $\pi'$.

After analyzing sustainable conditions in the second-stage, we then step back to analyze sustainable conditions in the first stage when both users choose their strategies in the first-stage before knowing their productivities. In the first stage, users decide the optimal chosen rate $\lambda$, and follow the recommended strategy $\pi$ in their self-interest. Under the service exchange dilemma game, a $\theta$-user will find it optimal to choose to be a client in the first stage, as his revenue is maximized when his matched $\widetilde\theta$-server chooses to follow the recommended strategy $\pi$ in the second stage, which yields payoff $b_{\pi}(\widetilde\theta, \theta)$ for him. On the contrary, choosing to be a sever will suffer a cost $c_{\pi}(\theta, \widetilde\theta)$. However, social welfare is maximized if and only if every user chooses to be a server or a client with the same probability $\lambda=\frac{1}{2}$, which we name it as the principle of fairness inspired by \cite{partov}, and derive incentive constraints that characterize sustainable conditions in the first stage as shown in Lemma 2.

\begin{myLem}
(\textbf{The Principle of Fairness}) A two-sided rating protocol $\mathcal{P}$ satisfies the principle of fairness if and only if
\begin{equation} {}
  \label{tpof}
  \begin{split}
 &\frac{\frac{1}{2}[\eta_\mathcal{P}(1)(1-2\varepsilon)b+\eta_\mathcal{P}(0)s]}{1+\omega(\phi_1-\phi_0)}\geq \\ &~~~~\max\Big\{\frac{(1-\varepsilon)b-c+\eta_\mathcal{P}(1)s}{\omega[(1-\varepsilon)\alpha_1+\varepsilon(1-\beta_1)+\gamma_1-2(1-\delta_1)]},\\ &~~~~~~~~~~~\frac{[\eta_\mathcal{P}(0)(1-\varepsilon)+\eta_\mathcal{P}(1)\varepsilon]b-c+s}{\omega[(1-\varepsilon)\alpha_0+\varepsilon(1-\beta_0)+\gamma_0-2(1-\delta_0)]}\Big\}
 \end{split}
 \end{equation}
\end{myLem}

\begin{IEEEproof}
For the ``if'' part: Assume that each user selects $\lambda=\frac{1}{2}$ in the first stage, and adopts the recommend strategy $\pi$ in the second stage, then his expected long-term utility can be expressed as
\begin{equation} {}
\label{gpyz0}
  \begin{split}
v_{\mathcal{P},\lambda=\frac{1}{2}}^\infty (\theta)= &v_{\mathcal{P},\lambda=\frac{1}{2}} (\theta) \\&+\omega \sum_{\theta' \in \Theta} p_{\mathcal{P},\lambda=\frac{1}{2}}(\theta'|\theta,\alpha)v_{\mathcal{P},\lambda=\frac{1}{2}}^\infty (\theta')
 \end{split}
 \end{equation}
Where $p_{\mathcal{P},\lambda=\frac{1}{2}}(\theta'|\theta)$ is the transition probability that a compliant $\theta$-user becomes a $\theta'$-user in the next period when he selects $\lambda=\frac{1}{2}$ in the first stage under the rating protocol $\mathcal{P}$, which can be found in Eq.(\ref{two}).

As a $\theta$-user can receive the benefit $b_{\pi}(\theta, \widetilde\theta)$ if and only if he chooses to be a client in the current period under the recommended strategy $\pi$, otherwise, he will suffer a cost $c_{\pi}(\theta_S, \theta_C)$. \textcolor{black}{We now suppose that a user deviates from $\lambda=\frac{1}{2}$ to $\lambda'\neq \frac{1}{2}$ in the current period, and follows $\lambda=\frac{1}{2}$ afterwards. It is obvious that $\lambda'>\frac{1}{2}$, because rewards for $\lambda=\frac{1}{2}$ and $\lambda'<\frac{1}{2}$ are the same, while a higher cost will be suffered by selecting to be a server with a higher probability $1-\lambda'$. Assuming that $\frac{1}{2}<\lambda_1'<\lambda_2'\leq 1$, and according to Eq.(\ref{lpv}), we have}
\begin{equation} {}
\label{chazhi}
  \begin{split}
&\textcolor{black}{v_{\mathcal{P},\lambda_2'}^\infty (\theta)-v_{\mathcal{P},\lambda_1'}^\infty (\theta)=(v_{\mathcal{P},\lambda_2'}(\theta)-v_{\mathcal{P},\lambda_2'}(\theta))}\\
&~~~~\textcolor{black}{+\omega \big\{v_{\mathcal{P},\lambda=\frac{1}{2}}^\infty (1)[p_{\mathcal{P},\lambda_2'}(1|\theta)-p_{\mathcal{P},\lambda_1'}(1|\theta)]}\\
&~~~~~~~~\textcolor{black}{+v_{\mathcal{P},\lambda=\frac{1}{2}}^\infty (0)[p_{\mathcal{P},\lambda_2'}(0|\theta)-p_{\mathcal{P},\lambda_1'}(0|\theta)]\big\}}
 \end{split}
 \end{equation}

\textcolor{black}{According to Eq.(\ref{two}), we have}
\begin{equation} {}
\label{gailv1}
  \begin{split}
 &\textcolor{black}{p_{\mathcal{P},\lambda_2'}(1|\theta)-p_{\mathcal{P},\lambda_1'}(1|\theta)}\\
 &~~~~\textcolor{black}{=(\lambda_2'-\lambda_1')[\gamma_{\theta}-(1-\varepsilon)\alpha_{\theta}-\varepsilon(1-\beta_\theta)]}
 \end{split}
 \end{equation}
\begin{equation} {}
\label{gailv2}
  \begin{split}
 &\textcolor{black}{p_{\mathcal{P},\lambda_2'}(0|\theta)-p_{\mathcal{P},\lambda_1'}(0|\theta)}\\
 &~~~~\textcolor{black}{=(\lambda_2'-\lambda_1')[(1-\gamma_\theta)-(1-\varepsilon)(1-\alpha_\theta)-\varepsilon\beta_\theta]}\\
 &~~~~\textcolor{black}{=(\lambda_2'-\lambda_1')[\gamma_{\theta}-(1-\varepsilon)\alpha_{\theta}-\varepsilon(1-\beta_\theta)]}
 \end{split}
 \end{equation}

\textcolor{black}{Hence, Eq.(\ref{chazhi}) can be rewritten as}
\begin{equation} {}
\label{chazhi2}
  \begin{split}
&\textcolor{black}{v_{\mathcal{P},\lambda_2'}^\infty (\theta)-v_{\mathcal{P},\lambda_1'}^\infty (\theta)=(v_{\mathcal{P},\lambda_2'}(\theta)-v_{\mathcal{P},\lambda_2'}(\theta))} \\
&~~~~\textcolor{black}{+\omega(\lambda_2'-\lambda_1')[\gamma_{\theta}-(1-\varepsilon)\alpha_{\theta}-\varepsilon(1-\beta_\theta)]}\\
&~~~~~~~~\textcolor{black}{(v_{\mathcal{P},\lambda=\frac{1}{2}}^\infty (1)-v_{\mathcal{P},\lambda=\frac{1}{2}}^\infty (0))}
 \end{split}
 \end{equation}

\textcolor{black}{We can derive that $v_{\mathcal{P},\lambda_2'}^\infty (\theta)-v_{\mathcal{P},\lambda_1'}^\infty (\theta)$ is a constant according to Eq.(\ref{bp20}), that is $v_{\mathcal{P},\lambda'}^\infty (\theta),\forall\lambda'\in[\frac{1}{2},1]$ is a monotonic function, which is determined by the intrinsic parameters (\emph{i.e.}, $b$, $c$, $s$, $\varepsilon$ and $\omega$), as well as the design parameters (\emph{i.e.}, $(\alpha_\theta^\ast,\beta_\theta^\ast,\gamma_\theta^\ast,\delta_\theta^\ast), \forall \theta\in\Theta$). Assuming that $v_{\mathcal{P},\lambda'}^\infty (\theta),\forall\lambda'\in[\frac{1}{2},1]$ is monotonic decreasing with $\lambda'$, then no user will deviate from $\lambda=\frac{1}{2}$ as it is his optimal choice. Therefore, we only need to check the case that $v_{\mathcal{P},\lambda'}^\infty (\theta),\forall\lambda'\in[\frac{1}{2},1]$ is monotonic increasing with $\lambda'$. It is obvious that the expected long-term utility of a user has its maximum value at $\lambda'$=1. Without loss of generality, we now suppose that a user deviates from $\lambda=\frac{1}{2}$ to $\lambda'$=1 in the current period, and follows $\lambda=\frac{1}{2}$ afterwards, then his expected long-term utilities can be expressed as}
\begin{equation} {}
\label{gpyz1}
  \begin{split}
v_{\mathcal{P},\lambda=1}^\infty (\theta)= &v_{\mathcal{P},\lambda=1} (\theta)+\omega \sum_{\theta' \in \Theta} p_{\mathcal{P},\lambda=1}(\theta'|\theta)v_{\mathcal{P},\lambda=\frac{1}{2}}^\infty (\theta')
 \end{split}
 \end{equation}
Where $p_{\mathcal{P},\lambda=1}(\theta'|\theta)$ can be computed based on Eq.(\ref{updatec}).
\begin{eqnarray}  {p_{\mathcal{P},\lambda=1}(\theta'|\theta)=}
\label{two2}
  \left \{
\begin{array}{ll}
      1-\delta_{\theta}, &if \theta'=1\\
      \delta_{\theta}, &if \theta'=0\\
\end{array}
\right.
\end{eqnarray}
By comparing Eq.(\ref{gpyz0}) with Eq.(\ref{gpyz1}), we can check whether a $\theta$-user has an incentive to deviate from $\lambda=\frac{1}{2}$ as follows
	
\begin{equation} {}
\begin{split}
\label{com20}
&v_{\mathcal{P},\lambda=\frac{1}{2}}^\infty (\theta)-v_{\mathcal{P},\lambda=1}^\infty (\theta)= v_{\mathcal{P},\lambda=\frac{1}{2}}(\theta)-v_{\mathcal{P},\lambda=1} (\theta)+\\
&~~~~~~~~~~\frac{\omega}{2}{[(1-\varepsilon)\alpha_\theta+\varepsilon(1-\beta_\theta)+\gamma_\theta-2(1-\delta_\theta)]}\\
&~~~~~~~~~~~~~~~~~~~~~~~~~~(v_{\mathcal{P},\lambda=\frac{1}{2}}^\infty (1)-v_{\mathcal{P},\lambda=\frac{1}{2}}^\infty (0))\geq 0
\end{split}
\end{equation}
With simple manipulation based on Eq.(\ref{pp}), we can obtain that $v_{\mathcal{P},\lambda=1}(1)=(1-\varepsilon)b-c$ and $v_{\mathcal{P},\lambda=1}(0)=[\eta_\mathcal{P}(0)(1-\varepsilon)+\eta_\mathcal{P}(1)\varepsilon]b-c$, together with Eq.(\ref{bp20}), we have
\begin{equation} {}
\begin{split}
\label{cz1}
&v_{\mathcal{P},\lambda=1}(1)-v_{\mathcal{P},\lambda=\frac{1}{2}}(1)=\frac{1}{2}[(1-\varepsilon)b-c+\eta_\mathcal{P}(1)s]
\end{split}
\end{equation}
\begin{equation} {}
\begin{split}
\label{cz2}
&v_{\mathcal{P},\lambda=1}(0)-v_{\mathcal{P},\lambda=\frac{1}{2}}(0)=\\
&~~~~~~~~~~~~~~~~~\frac{1}{2}\big\{[\eta_\mathcal{P}(0)(1-\varepsilon)+\eta_\mathcal{P}(1)\varepsilon]b-c+s\big\}
\end{split}
\end{equation}
If $\theta$=1, Eq.(\ref{com20}) can be rewritten as
\begin{equation} {}
  \label{bp1}
  \begin{split}
&v_{\mathcal{P},\lambda=\frac{1}{2}}^\infty (1)-v_{\mathcal{P},\lambda=\frac{1}{2}}^\infty (0)\geq \\ &~~~~~~~~\frac{(1-\varepsilon)b-c+\eta_\mathcal{P}(1)s}{\omega[(1-\varepsilon)\alpha_1+\varepsilon(1-\beta_1)+\gamma_1-2(1-\delta_1)]}
 \end{split}
 \end{equation}
While if $\theta$=0, Eq.(\ref{com20}) can be rewritten as
\begin{equation} {}
  \label{bp2}
  \begin{split}
&v_{\mathcal{P},\lambda=\frac{1}{2}}^\infty (1)-v_{\mathcal{P},\lambda=\frac{1}{2}}^\infty (0)\geq\\ &~~~~~~~~\frac{[\eta_\mathcal{P}(0)(1-\varepsilon)+\eta_\mathcal{P}(1)\varepsilon]b-c+s}{\omega[(1-\varepsilon)\alpha_0+\varepsilon(1-\beta_0)+\gamma_0-2(1-\delta_0)]}
 \end{split}
 \end{equation}

Combining Eq.(\ref{bp1}) and Eq.(\ref{bp2}), sufficient conditions that a two-sided rating protocol $\mathcal{P}$ is satisfied with the principle of fairness can be obtained, as shown in inequality (\ref{tpof}).

For the ``only if'' part: Suppose $\mathcal{P}$ is satisfied with the principle of fairness, then clearly there are no profitable deviations (\emph{i.e.}, $\rho>1$ or $\lambda>\frac{1}{2}$) in the first stage. We can prove the converse by showing that if $\mathcal{P}$ is not satisfied with the principle of fairness, there is at least one profitable deviation. Since the RHS of Eq.(\ref{tpof}) is bounded, this is true by the unimprovability property in Markov decision theory.
\end{IEEEproof}

Using one-shot deviation principle and the principle of fairness, we can derive incentive constraints that characterize necessary and sufficient conditions for a two-sided rating protocol to be sustainable, which is formalized in the next theorem.

\begin{myThe}{}
  A two-sided rating protocol $\mathcal{P}$ is sustainable if and only if it is satisfied with both of the one-shot deviation principle and the principle of fairness.
\end{myThe}

    \begin{IEEEproof}{}
    This proof can be directly obtained from Lemma 1 and 2, and is omitted here.
    \end{IEEEproof}

\subsection{Optimization Problem with Constraints}
\textcolor{black}{Under a sustainable rating protocol $\mathcal{P}$, it is in the self-interest of each user to devote a high level of effort (\emph{i.e.}, the one-shot deviation principle) and take the incentive to serve others (\emph{i.e.}, the principle of fairness). Obviously, a sustainable two-sided rating protocol always achieves a higher social welfare than a non-sustainable one, and hence it is only necessary to consider sustainable protocols in order to maximize the social welfare. We assume that the protocol designer is profit-seeking and aims to design a rating protocol $\mathcal{P}$ that maximizes the expected one-period utility a user obtains in one transaction, which is defined as the social welfare $U_\mathcal{P}$ in this paper.} As a result, the design of the two-sided rating protocol that maximizes the social welfare can be formulated as follows:

\begin{myDef} The two-sided rating protocol design problem can be formulated as follows:
      \begin{equation} {}
  \label{problem}
  \left\{
  \begin{aligned}
  \begin{split}
&\mathop{\max}\limits_{(\tau,\pi)}U_P \triangleq \sum_{\theta \in \Theta }\eta_P (\theta) v_P (\theta)=\frac{1}{2}\Big\{\eta_\mathcal{P}^2(0)\big\lbrack(1-2\varepsilon)b-s\big\rbrack-\\
&~~~~~\eta_\mathcal{P}(0)\big\lbrack(1-2\varepsilon)b-s\big\rbrack+(1-\varepsilon)b-c-s\Big\}\\
&s.t.~\frac{\frac{1}{2}[\eta_\mathcal{P}(1)(1-2\varepsilon)b+\eta_\mathcal{P}(0)s]}{1+\omega(\phi_1-\phi_0)}\geq \\
&~~~~~~\max\Big\{\frac{s}{\omega(1-2\varepsilon)(\alpha_0+\beta_0-1)},\\
&~~~~~~\frac{\eta_{\mathcal{P}}(1)s}{\omega(1-2\varepsilon)(\alpha_1+\beta_1-1)},\\
&~~~~~~\frac{(1-\varepsilon)b-c+\eta_\mathcal{P}(1)s}{\omega[(1-\varepsilon)\alpha_1+\varepsilon(1-\beta_1)+\gamma_1-2(1-\delta_1)]},\\ &~~~~~~\frac{[\eta_\mathcal{P}(0)(1-\varepsilon)+\eta_\mathcal{P}(1)\varepsilon]b-c+s}{\omega[(1-\varepsilon)\alpha_0+\varepsilon(1-\beta_0)+\gamma_0-2(1-\delta_0)]}\Big\}
 \end{split}
 \end{aligned}
  \right.
  \end{equation}
\end{myDef}

\textcolor{black}{It should be noted that both $\alpha_0+\beta_0\neq 1$ and $\alpha_1+\beta_1\neq 1$, otherwise Eq.(\ref{updates}) will be rewritten as}
\begin{eqnarray}{\textcolor{black}{\tau_S (\theta'_S| \theta_S, \theta_C, r, \pi)=}}
\label{updatess}
      \left \{
\begin{array}{ll}
  \textcolor{black}{\alpha_{\theta_S},}  &\textcolor{black}{\theta'_S=1}\\
  \textcolor{black}{1-\alpha_{\theta_S},}  &\textcolor{black}{\theta'_S=0}
\end{array}
\right.
\end{eqnarray}
\textcolor{black}{which is independent of users' behaviors, and thus cannot provide effective incentives.}

\section{OPTIMAL DESIGN OF Two-Sided RATING PROTOCOLS}

In this section we investigate the design of an optimal two-sided rating protocol that solves the two-sided rating protocol design problem under a given recommended strategy $\pi$, \emph{i.e.}, selecting the optimal rating update rule $\tau$, which are determined by design parameters $(\alpha_\theta,\beta_\theta,\gamma_\theta,\delta_\theta), \forall \theta\in\Theta$. In order to characterize an optimal design which is denoted as $(\alpha_\theta^\ast,\beta_\theta^\ast,\gamma_\theta^\ast,\delta_\theta^\ast), \forall \theta\in\Theta$, we investigate the impacts of design parameters on the social welfare $U_\mathcal{P} \triangleq \sum_{\theta \in \Theta }\eta_P (\theta) v_P (\theta)$, and the incentive for satisfying constraints in Eq.(\ref{problem}).
\subsection{Existence of a Sustainable Two-sided Rating Protocol}
We first investigate whether there exists a sustainable two-sided rating protocol under $\pi$,\emph{ i.e.}, determining whether there exists a feasible solution for the design problem of Eq.(\ref{problem}).
\begin{myThe}{}
  A sustainable two-sided rating protocol $\mathcal{P}$ under the recommended strategy $\pi$ exists if and only if
  \begin{equation} {}
  \label{existence}
  \begin{split}
&\omega \geq \max\Big\{\frac{2s}{(1-2\varepsilon)[(1-\frac{5}{2}\varepsilon+\varepsilon^2)b+\frac{\varepsilon}{2}s]},\\
&\frac{(2-2\varepsilon)b-2c+(2-\varepsilon)s}{[(1-\frac{5}{2}\varepsilon+\varepsilon^2)b+\frac{\varepsilon}{2}s](2-\varepsilon)},\frac{(3\varepsilon-2\varepsilon^2)b-2c+2s}{[(1-\frac{5}{2}\varepsilon+\varepsilon^2)b+\frac{\varepsilon}{2}s](2-\varepsilon)}\Big\}
 \end{split}
  \end{equation}
\end{myThe}

    \begin{IEEEproof}{}
For the ``if'' part: Among the eight design parameters, $\alpha_1$, $\alpha_0$, $\gamma_1$ and $\gamma_0$ can be referred to as reward factors imposed on compliant users, while $\beta_1$, $\beta_0$, $\delta_1$ and $\delta_0$ can be referred to as punishment factors imposed on non-compliant users. The incentive for self-interested users to be a compliant user is maximized when we maximize all of reward factors and punishment factors, \emph{i.e.}, $\alpha_\theta=\beta_\theta=\gamma_\theta=\delta_\theta=1,\forall\theta\in\Theta$. Then, Eq.(\ref{problem}) can be transformed into

\begin{equation} {}
  \label{problem01}
  \begin{split}
&\frac{1}{2}[(1-\frac{\varepsilon}{2})(1-2\varepsilon)b+\frac{\varepsilon}{2}s]\geq \max\Big\{\frac{s}{\omega(1-2\varepsilon)},\frac{(1-\frac{\varepsilon}{2})s}{\omega(1-2\varepsilon)},\\
&~~~~~~~~~~~~~~~\frac{(1-\varepsilon)b-c+(1-\frac{\varepsilon}{2})s}{\omega(2-\varepsilon)}, \frac{(\frac{3}{2}\varepsilon-\varepsilon^2)b-c+s}{\omega(2-\varepsilon)}\Big\}
 \end{split}
  \end{equation}

It is obvious that $\frac{s}{\omega(1-2\varepsilon)}>\frac{(1-\frac{\varepsilon}{2})s}{\omega(1-2\varepsilon)}$, hence, Eq.(\ref{problem01}) can be revised as follows
\begin{equation} {}
  \label{problem02}
  \begin{split}
&\frac{1}{2}[(1-\frac{5}{2}\varepsilon+\varepsilon^2)b+\frac{\varepsilon}{2}s]\geq \max\Big\{\frac{s}{\omega(1-2\varepsilon)},\\
&~~~~~~~~~\frac{(1-\varepsilon)b-c+(1-\frac{\varepsilon}{2})s}{\omega(2-\varepsilon)}, \frac{(\frac{3}{2}\varepsilon-\varepsilon^2)b-c+s}{\omega(2-\varepsilon)}\Big\}
 \end{split}
  \end{equation}

By solving Eq.(\ref{problem02}), we can obtain Eq.(\ref{existence}), that is a sufficient condition that there exists a feasible solution for the design problem. It shows that Eq.(\ref{problem}) always has a feasible solution if users have sufficient patience (\emph{i.e.}, when the discount factor $\omega$ is large). \textcolor{black}{Therefore, there exists a sustainable two-sided rating protocol if Eq.(\ref{existence}) holds, and $\omega<1$ as no one can be 100\% patient.}

For the ``only if'' part: Suppose Eq.(\ref{existence}) hold, it is easy to determine whether constraints in the design problem of (\ref{problem}) are satisfied, similar as the above, the ``only if'' part can be proved.
\end{IEEEproof}

\subsection{Optimal Values of the Rating Update Rule}
In this section, we assume that Eq.(\ref{existence}) holds, that is, there exists a feasible solution for the two-sided rating protocol design problem of Eq.(\ref{problem}). Our goal is to select $(\alpha_\theta^\ast,\beta_\theta^\ast,\gamma_\theta^\ast,\delta_\theta^\ast), \forall \theta\in\Theta$ to maximize the social welfare $U_\mathcal{P}$, that is, maximizing reward factors $\alpha_\theta,\gamma_\theta, \forall \theta\in\Theta$, and minimizing punishment factors $\beta_\theta$ and $\delta_\theta, \forall \theta\in\Theta$. With this idea, Theorem 3 gives the optimal value of reward/punishment factors except $\beta_\theta,\forall \theta\in\Theta$.

\begin{myThe}{}
Given a sustainable two-sided rating protocol $\mathcal{P}$,  $\alpha_\theta^\ast=\gamma_\theta^\ast=\delta_\theta^\ast=1, \forall \theta\in\Theta$ is always the optimal solution to Eq.(\ref{problem}).
\end{myThe}
\begin{IEEEproof}{}
\textcolor{black}{Social welfare $U_\mathcal{P}$ is monotone decreasing with $\eta_\mathcal{P} (0)\in[0,1]$ according to Eq.(\ref{problem}), where $\eta_\mathcal{P}$ is monotone decreasing with $\alpha_\theta\in[0,1]$ and $\beta_\theta\in[0,1]$ for all $\theta\in\Theta$ according to Eq.(\ref{a1}), and thus $U_\mathcal{P}$ is monotone increasing with reward factors $\alpha_\theta,\gamma_\theta, \forall \theta\in\Theta$, and the upper bound of them is 1, with which the incentive constraints in Eq.(\ref{problem}) are satisfied.} As $U_\mathcal{P}$ monotonically decreases with punishment factors $\beta_\theta,\forall \theta\in\Theta$, and given $\alpha_\theta^\ast=\gamma_\theta^\ast=1, \forall \theta\in\Theta$, the design problem is transformed into the selection of the smallest $\beta_\theta,\forall \theta\in\Theta$, with which the incentive constraints in Eq.(\ref{problem}) are satisfied. It is obvious that the smallest $\beta_\theta,\forall \theta\in\Theta$ can be obtained when we select the largest $\delta_\theta,\forall \theta\in\Theta$, and the upper bound of $\delta_\theta,\forall \theta\in\Theta$ is 1. Since  $U_\mathcal{P}$ is only determined by $\beta_\theta,\forall \theta\in\Theta$, rather than $\delta_\theta,\forall \theta\in\Theta$, in order to provide sufficient incentive and get as less $\beta_\theta,\forall \theta\in\Theta$ as possible, we have $\delta_\theta=1,\forall \theta\in\Theta$. Hence, this statement follows.
\end{IEEEproof}

By substituting $\alpha_\theta=\gamma_\theta=\delta_\theta=1, \forall \theta\in\Theta$ into Eq.(\ref{a1}), we have
  \begin{equation} {}
  \label{eta2}
  \left\{
  \begin{aligned}
     & \eta_\mathcal{P} (0)=\frac{\varepsilon\beta_1}{2+\varepsilon\beta_1-\varepsilon\beta_0}\\
     & \eta_\mathcal{P} (1)=\frac{2-\varepsilon\beta_0}{2+\varepsilon\beta_1-\varepsilon\beta_0}\\
 \end{aligned}
  \right.
  \end{equation}
 It is obvious that $0<\eta_\mathcal{P} (0)<\frac{1}{2}<\eta_\mathcal{P} (1)$ as $\beta_1,\beta_0\in[0,1]$. \textcolor{black}{Since the social welfare $U_\mathcal{P}$ is monotone decreasing with $\eta_\mathcal{P} (0)$, it can be shown that the problem of maximizing $U_\mathcal{P}$ is equivalent to the problem of minimizing $\eta_\mathcal{P} (0)$. By substituting $\alpha_\theta=\gamma_\theta=\delta_\theta=1, \forall \theta\in\Theta$ into the constraints in Eq.(\ref{problem}), and replacing the objective function $U_\mathcal{P}$ with $\eta_\mathcal{P} (0)$, the design problem w.r.t $\beta_0$ and $\beta_1$ can be rewritten as}
\begin{equation} {}
  \label{problem10}
  \left\{
  \begin{aligned}
  \begin{split}
&\mathop{\min}\limits_{(\beta_0,\beta_1)}~\frac{1}{1+\frac{2-\varepsilon\beta_0}{\varepsilon\beta_1}}\\
&s.t.~\frac{(2-\varepsilon\beta_0)(1-2\varepsilon)b+s\varepsilon\beta_1}{(2+\varepsilon\beta_1-\varepsilon\beta_0)[2+\omega\varepsilon(\beta_1-\beta_0)]}\geq \\
&~~~~\max\Big\{\frac{s}{\omega(1-2\varepsilon)\beta_0},~\frac{s(2-\varepsilon\beta_0)}{\omega(1-2\varepsilon)(2+\varepsilon\beta_1-\varepsilon\beta_0)\beta_1},\\
&~~~~\frac{[(1-\varepsilon)b-c](2+\varepsilon\beta_1-\varepsilon\beta_0)+s(2-\varepsilon\beta_0)}{\omega(2+\varepsilon\beta_1-\varepsilon\beta_0)(2-\varepsilon\beta_1)},\\ &~~~~\frac{[(1-\varepsilon)\varepsilon\beta_1+\varepsilon(2-\varepsilon\beta_0)]b+(s-c)(2+\varepsilon\beta_1-\varepsilon\beta_0)}{\omega(2+\varepsilon\beta_1-\varepsilon\beta_0)(2-\varepsilon\beta_0)}\Big\}
 \end{split}
 \end{aligned}
  \right.
  \end{equation}

\textcolor{black}{It is obvious that Eq.(\ref{problem10}) is a non-convex optimization problem as the second constraint is a non-convex inequality (It should be noted that Eq.(\ref{problem10}) consists of four constraint inequalities, which is written compactly just to save space.). We now design an algorithm to this problem inspired by \cite{liu1,liu2}, it achieves low-complexity computation via a two-stage two-step alternate manner.} The proposed algorithm is outlined in Algorithm 1, where $obj^t$ denotes value of $\frac{1}{1+\frac{2-\varepsilon\beta_0}{\varepsilon\beta_1}}$ at the \emph{t}-th iteration. It should be noted that the value of $obj^t$ of Algorithm 1 is guaranteed to be monotonically decreasing when optimizing one variable with another fixed in each iteration \cite{bezdek}. Meanwhile, $\frac{1}{1+\frac{2-\varepsilon\beta_0}{\varepsilon\beta_1}}$ is lower-bounded by $\frac{\varepsilon}{2}$ with the presence of imperfect monitoring (\emph{i.e.}, $\varepsilon>0$). Therefore, Algorithm 1 is guaranteed to converge. The detailed explanation of Algorithm 1 can be found in the proof of Theorem 4.

\begin{algorithm}[bt]
\caption{Alternate Optimal Design of Punishment factors $\beta_0$ and $\beta_1$}
\label{alg1}
\begin{algorithmic}[1]
\REQUIRE $b$, $c$, $s$, $\varepsilon$, $\omega$ and $\epsilon$.
\ENSURE $\beta_0^\ast$ and $\beta_1^\ast$.
\STATE Initialize $\beta_0^0=1$ and $t=1$.
\REPEAT
\STATE Update $\beta_1^t$ by solving Eq.(\ref{beta110}) with given $\beta_0^{t-1}$.
\STATE Update $\beta_0^t$ by solving Eq.(\ref{beta001}) with given $\beta_1^{t}$.
\STATE $t=t+1$.
\UNTIL $(obj^{t-1}-obj^t)/{obj^t}\leq\epsilon$
\STATE $obj^\ast=obj^t$
\STATE Set $\beta_1^0=1$ and $t=1$.
\REPEAT
\STATE Update $\beta_0^t$ by solving Eq.(\ref{beta001}) with given $\beta_1^{t-1}$.
\STATE Update $\beta_1^t$ by solving Eq.(\ref{beta110}) with given $\beta_0^{t}$.
\STATE $t=t+1$.
\UNTIL $(obj^{t-1}-obj^t)/{obj^t}\leq\epsilon$
\STATE $(\beta_0^\ast,\beta_1^\ast)=\mathop{\arg}\limits_{(\beta_0,\beta_1)}\min\{obj^\ast,obj^t\}$.
\end{algorithmic}
\end{algorithm}

\begin{myThe}{}
Given $\alpha_\theta^\ast=\gamma_\theta^\ast=\delta_\theta^\ast=1, \forall \theta\in\Theta$, and a residual $\epsilon$, the output of $\beta_\theta^\ast, \forall \theta\in\Theta$ by Algorithm 1 is an optimal solution to Eq.(\ref{problem10}).
\end{myThe}
\begin{IEEEproof}{}
Algorithm 1 consists of two stages, and each stage consists of two steps. In stage (\emph{\romannumeral1}), we first fix $\beta_0=1$, and then update both $\beta_1$ and $\beta_0$. Where stage (\emph{\romannumeral2}) is symmetric with stage (\emph{\romannumeral1}), the only difference is that we first fix $\beta_1=1$ and then update both $\beta_0$ and $\beta_1$.

\textbf{Step (\emph{\romannumeral1}): Optimizing with fixed} $\bm{\beta_0}$. Given $\beta_0$, the optimization problem in Eq.(\ref{problem10}) w.r.t $\beta_1$ can be rewritten as
\begin{equation} {}
  \label{beta188}
  \left\{
  \begin{array}{ll}
&\mathop{\min} \beta_1\\
&s.t.
\left\{
\begin{array}{ll}x_i\beta_1^2+y_i\beta_1+z_i\leq 0, i\in\{1,3,4\}\\
x_j\beta_1^2+y_j\beta_1+z_j\geq 0, j\in\{2\} \end{array}\right.\\
\end{array}
\right.
\end{equation}
Where $x_1=s\varepsilon^2\omega$, $y_1=s\omega(2+2\omega-\omega\beta_0)$, $z_1=(2-\varepsilon\beta_0)[s(2-\omega\varepsilon\beta_0)-\omega b(1-2\varepsilon)^2\beta_0]$, $x_2=s\varepsilon\omega(1-2\varepsilon)$, $y_2=\omega(2-\varepsilon\beta_0)[b(1-2\varepsilon)^2-s\varepsilon]$, $z_2=-s(2-\omega\varepsilon\beta_0)(2-\varepsilon\beta_0)$, $x_3=\varepsilon^2\omega(b-b\varepsilon-c+s)$, $y_3=\varepsilon\omega(2-\varepsilon\beta_0)(s-c+2b-3b\varepsilon)+\varepsilon(b-b\varepsilon-c)(2-\omega\varepsilon\beta_0)-2s\varepsilon\omega$, $z_3=(2-\varepsilon\beta_0)[(2-\varepsilon\omega\beta_0)(s+b-b\varepsilon-c)-2b\omega(1-2\varepsilon)]$, $x_4=\varepsilon^2\omega(b-b\varepsilon+s-c)$, $y_4=\varepsilon\omega(2-\varepsilon\beta_0)(b\varepsilon-c)+\varepsilon(2-\varepsilon\omega\beta_0)(s-c+b-b\varepsilon)$, $z_4=(2-\varepsilon\beta_0)[(2-\omega\varepsilon\beta_0)(s-c+b\varepsilon)-b\omega(1-2\varepsilon)(2-\varepsilon\beta_0)]$.

 By solving inequalities in Eq.(\ref{beta188}), we have
  \begin{equation} {}
  \label{beta111}
  \left\{
  \begin{aligned}
     &\beta_1\in \Big\lbrack-\frac{y_i}{2x_i}-\sqrt{(\frac{y_i}{2x_i})^2-z_i}, -\frac{y_i}{2x_i}+\sqrt{(\frac{y_i}{2x_i})^2-z_i}\Big\rbrack,\\
     &~~~~~~~~~~~~~~~~~~~~~~~~~~~~~~~~~~~~~~~~~~~~~~~~~~~\forall i\in\{1,3,4\}\\
     &\beta_1\in \Big\lbrack 0,\frac{y_2}{2x_2}-\sqrt{(\frac{y_2}{2x_2})^2-z_2}\Big\rbrack\cup \\
     &~~~~~~~~~~~~~~~~~~~~~~~~~~~~~~~~~~~~~\Big\lbrack\frac{y_2}{2x_2}+\sqrt{(\frac{y_2}{2x_2})^2-z_2}, 1\Big\rbrack\\
 \end{aligned}
  \right.
  \end{equation}

Let $\psi_u^1=\frac{y_2}{2x_2}-\sqrt{(\frac{y_2}{2x_2})^2-z_2}$, $\psi_u^2=\mathop{\min}\limits_{i\in{\{1,3,4\}}}\Big\{-\frac{y_i}{2x_i}+\sqrt{(\frac{y_i}{2x_i})^2-z_i}\Big\}$, $\psi_l^1=\mathop{\max}\limits_{i\in{\{1,3,4\}}}\Big\{-\frac{y_i}{2x_i}-\sqrt{(\frac{y_i}{2x_i})^2-z_i}\Big\}$, $\psi_l^2=\frac{y_2}{2x_2}+\sqrt{(\frac{y_2}{2x_2})^2-z_2}$, $\Psi_1=[\psi_l^1,\psi_u^1]$, and $\Psi_2=[\psi_l^2,\psi_u^2]$, \textcolor{black}{then Eq.(\ref{beta111}) can be rewritten as}
 \begin{equation} {}
  \label{beta1880}
    \begin{split}
     \beta_1\in \Psi_1\cup\Psi_2
   \end{split}
   \end{equation}
The optimal value of $\beta_1$ for Eq.(\ref{beta188}) can be conducted as follows

\begin{equation} {}
  \label{beta110}
  \left\{
  \begin{aligned}
  \begin{split}
&\mathop{\min} \beta_1\\
&s.t.~\beta_1\in [0,1]\cap\mathop{\bigcap}\limits_{i\in\{1,2\}}\Psi_i
 \end{split}
 \end{aligned}
  \right.
  \end{equation}

\textbf{Step (\emph{\romannumeral2}): Optimizing with fixed} $\bm{\beta_1}$. \textcolor{black}{Given $\beta_1$, the optimization problem in Eq.(\ref{problem10}) w.r.t $\beta_0$ can be rewritten as}
\begin{equation} {}
  \label{beta088}
  \left\{
  \begin{array}{ll}
&\mathop{\min} \textcolor{black}{\beta_0}\\
&s.t. ~x_j'\beta_1^2+y_j'\beta_1+z_j'\leq 0, j\in\{1,2,3,4\}
\end{array}
\right.
\end{equation}
Where $x_1'=\varepsilon\omega[s\varepsilon+b(1-2\varepsilon)^2]$, $y_1'=-2b\omega(1-2\varepsilon)^2-s\varepsilon\omega(2+\varepsilon\beta_1)-s\varepsilon(2+\varepsilon\omega\beta_1)$, $z_1'=s(2+\varepsilon\beta_1)(2+\varepsilon\omega\beta_1)$, $x_2'=s\varepsilon^2\omega$, $y_2'=-s\varepsilon(2+\varepsilon\omega\beta_1)-2s\varepsilon\omega-b\varepsilon\omega(1-2\varepsilon)^2\beta_1$, $z_2'=2s(2+\omega\varepsilon\beta_1)-s\varepsilon\omega(1-2\varepsilon)\beta_1^2-2b\omega(1-2\varepsilon)^2\beta_1$, $x_3'=\varepsilon^2\omega(b-b\varepsilon-c+s)$, $y_3'=b\varepsilon\omega(1-2\varepsilon)(2-\varepsilon\beta_1)-\varepsilon(b-b\varepsilon-c)(2+2w+2w\varepsilon\beta_1)-s\varepsilon(2+\varepsilon\omega\beta_1)$, $z_3'=-w(2-\varepsilon\beta_1)[2b(1-2\varepsilon)+s\varepsilon\beta_1]+(2+\omega\varepsilon\beta_1)[(b-b\varepsilon-c)(2+\varepsilon\beta_1)+2s]$, $x_4'=\varepsilon^2\omega(3b\varepsilon+s-c-b)$, $y_4'=-\varepsilon\omega[b\varepsilon(\beta_1-\varepsilon\beta_1+10)+2s-2c-c\varepsilon\beta_1-4b]-\varepsilon(2+\varepsilon\omega\beta_1)(b\varepsilon+s-c)$, $z_4'=(2+\varepsilon\omega\beta_1)[b\varepsilon(1-\varepsilon)\beta_1+2b\varepsilon+(s-c)(2+\varepsilon\beta_1)]-2\omega[2b(1-2\varepsilon)+s\varepsilon\beta_1]$.

\textcolor{black}{Similarly, Eq.(\ref{beta088}) can be solved as follows}

 \begin{equation} {}
  \label{beta3}
    \begin{split}
     \beta_0\in\Big\lbrack-\frac{y_j'}{2x_j'}-\sqrt{(\frac{y_j'}{2x_j'})^2-z_j'},-\frac{y_j'}{2x_j'}+\sqrt{(\frac{y_j'}{2x_j'})^2-z_j'}\Big\rbrack,\\ \forall j\in\{1,2,3,4\}
   \end{split}
   \end{equation}

Let $\varphi_u=\mathop{\min}\limits_{j\in{\{1,2,3,4\}}}\Big\{-\frac{y_j'}{2x_j'}+\sqrt{(\frac{y_j'}{2x_j'})^2-z_j'}\Big\}$, and $\varphi_l=\mathop{\max}\limits_{j\in{\{1,2,3,4\}}}\Big\{-\frac{y_j'}{2x_j'}-\sqrt{(\frac{y_j'}{2x_j'})^2-z_j'}\Big\}$, \textcolor{black}{then Eq.(\ref{beta3}) can be rewritten as}

\begin{equation} {}
  \label{beta001}
  \left\{
  \begin{aligned}
  \begin{split}
&\mathop{\min} \beta_0\\
&s.t.~\beta_0\in [0,1]\cap[\varphi_l,\varphi_u]
 \end{split}
 \end{aligned}
  \right.
  \end{equation}

\textcolor{black}{Given $\beta_0=1$, we first calculate the smallest $\beta_1$ by solving Eq.(\ref{beta110}), with which the incentive constraints Eq.(\ref{beta188}) are satisfied, then update $\beta_0^t$ based on $\beta_1^t$, and next repeat lines 3, 4, and 5 until the termination condition in line 6 of Algorithm 1 is satisfied. Finally, we obtain a solution and denote it as ($\beta_0^{\romannumeral1},\beta_1^{\romannumeral1}$) based on step (\romannumeral1). Similarly, we can derive another solution ($\beta_0^{\romannumeral2},\beta_1^{\romannumeral2}$) based on step (\romannumeral2). And it is obvious that $\beta_0^{\romannumeral1}\ge\beta_0^{\romannumeral2}$ and $\beta_0^{\romannumeral1}\le\beta_0^{\romannumeral2}$.}

\textcolor{black}{We now assume that there exists another solution ($\beta_0^{\triangle},\beta_1^{\triangle}$) such that $obj^{\triangle}<\min\{obj^{\romannumeral1},obj^{\romannumeral2}\}$. Since Algorithm 1 will not stop to search a less value of $obj$ if $\beta_0^{\triangle}\not\in(\beta_0^{\romannumeral2},\beta_0^{\romannumeral1})$ or $\beta_1^{\triangle}\not\in(\beta_1^{\romannumeral1},\beta_1^{\romannumeral2})$, and hence we have $\beta_0^{\triangle}\in(\beta_0^{\romannumeral2},\beta_0^{\romannumeral1})$ and $\beta_1^{\triangle}\in(\beta_1^{\romannumeral1},\beta_1^{\romannumeral2})$. Without loss of generality, we assume that $obj^{\romannumeral1}<obj^{\romannumeral2}$, which means that ($\beta_0^{\romannumeral1},\beta_1^{\romannumeral1}$) is a better choice than ($\beta_0^{\romannumeral2},\beta_1^{\romannumeral2}$), and $\beta_1^{\romannumeral1}$ is an optimal solution. According to Algorithm 1, it updates $\beta_0^{t}$ by solving Eq.(\ref{beta001}) with given $\beta_0^{\triangle}>\beta_1^{\romannumeral1}$, and it will get a smaller value of $obj^{\triangle}$ by decreasing $\beta_1^{\triangle}$ to $\beta_1^{\romannumeral1}$, and thus ($\beta_0^{\triangle},\beta_1^{\triangle}$) is no better than ($\beta_0^{\romannumeral1},\beta_1^{\romannumeral1}$). In the case $obj^{\romannumeral1}>obj^{\romannumeral1}$, it can be proved that ($\beta_0^{\triangle},\beta_1^{\triangle}$) is no better than ($\beta_0^{\romannumeral2},\beta_1^{\romannumeral2}$) in a similar way, which is omitted here.}

\textcolor{black}{As a result, the output of Algorithm 1 is an optimal solution to Eq.(\ref{problem10}) with a given residual $\epsilon$.}
\end{IEEEproof}

\subsection{Optimal Values of the Rating Update Rules with a Stricter Recommended Strategy}

\begin{figure*}
\centering

\begin{minipage}{0.25\textwidth}
  \rightline{\includegraphics[width=\textwidth]{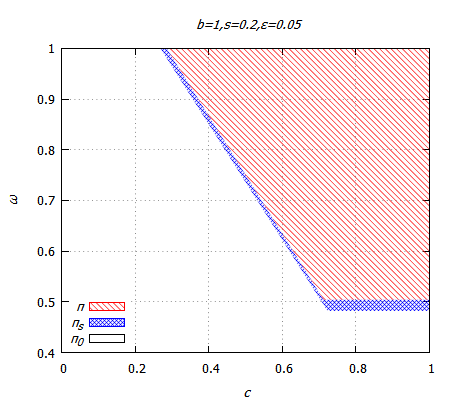}}
  \centerline{(\emph{a}) }
\end{minipage}
\begin{minipage}{0.25\textwidth}
  \leftline{\includegraphics[width=\textwidth]{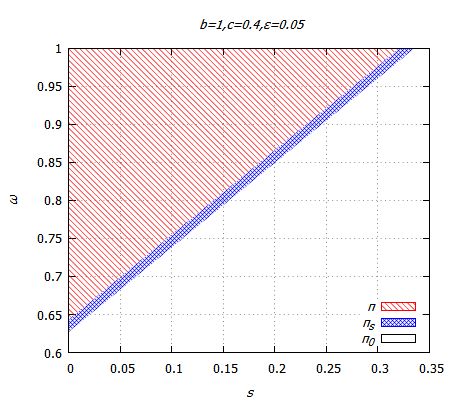}}
  \centerline{(\emph{b})}
\end{minipage}
\begin{minipage}{0.25\textwidth}
  \rightline{\includegraphics[width=\textwidth]{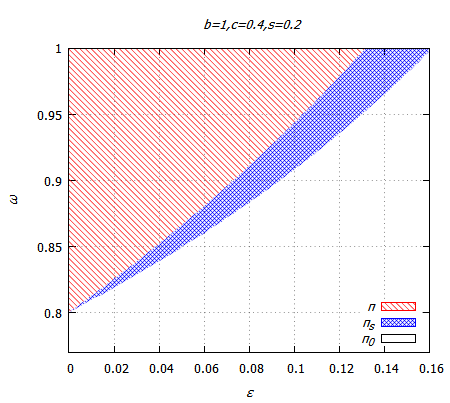}}
  \centerline{(\emph{c}) }
\end{minipage}
 \caption{Optimal recommended strategy versus intrinsic parameters: (a) $c$; (b) $s$; (c) $\varepsilon$.}
\label{fig2}
\end{figure*}

Eq.(\ref{existence}) shows that there exists a feasible solution for the design problem of Eq.(\ref{problem}) under the condition when the user is sufficiently patient with his discount factor $\omega$. However, such a condition may not hold with a small $\omega$ or $\max\{\frac{2s}{(1-2\varepsilon)[(1-\frac{5}{2}\varepsilon+\varepsilon^2)b+\frac{\varepsilon}{2}s]}$, $\frac{(2-2\varepsilon)b-2c+(2-\varepsilon)s}{[(1-\frac{5}{2}\varepsilon+\varepsilon^2)b+\frac{\varepsilon}{2}s](2-\varepsilon)}$, $\frac{(3\varepsilon-2\varepsilon^2)b-2c+2s}{[(1-\frac{5}{2}\varepsilon+\varepsilon^2)b+\frac{\varepsilon}{2}s](2-\varepsilon)}\}\geq 1$. In this condition, we should select a stricter recommended strategy $\pi_s$ among the total of $2^4$ possible recommended strategies, which is denoted as follows
 \begin{eqnarray}
 {\pi_s(\theta_S,\theta_C)=}
  \left \{
  \begin{array}{ll}
      1,  & if~ \theta_C\geq 1\\
      0,  & otherwise
  \end{array}
  \right.
\end{eqnarray}
Eq.(\ref{bp20}) will be rewritten as
\begin{equation} {}
  \label{bp201}
  \left\{
  \begin{aligned}
  \begin{split}
  &v_{\mathcal{P},\lambda,\pi_s}(0)=\lambda(\varepsilon b-c)-(1-\lambda)\eta_\mathcal{P}(1)s\\
  &v_{\mathcal{P},\lambda,\pi_s}(1)=\lambda[(1-\varepsilon)b-c]-(1-\lambda)\eta_\mathcal{P}(1)s\\
  &v_{\mathcal{P},\lambda,\pi_s}^\infty (1)-v_{\mathcal{P},\lambda,\pi_s}^\infty (0)=\frac{\lambda(1-2\varepsilon) b}{1+\omega(\phi_1-\phi_0)}
 \end{split}
 \end{aligned}
 \right.
 \end{equation}

The two-sided rating protocol design problem with a strict recommended strategy $\pi_s$, that is, Eq.(\ref{problem}) can be rewritten as follows:
      \begin{equation} {}
  \label{problem2}
  \left\{
  \begin{aligned}
  \begin{split}
&\mathop{\max}\limits_{(\tau,\pi_s)}U_P \triangleq \sum_{\theta \in \Theta }\eta_P (\theta) v_P (\theta)=\\ &~~~~~~\frac{1}{2}(1-\varepsilon)b-c-s-\eta_\mathcal{P}(0)[(1-2\varepsilon)b+s])\\
&s.t.~\frac{(1-2\varepsilon)b}{2+2\omega(\phi_1-\phi_0)}\geq \\
&~~~~~~\max\Big\{\frac{\eta_\mathcal{P}(1)s}{\omega(1-2\varepsilon)(\alpha_0+\beta_0-1)},\\
&~~~~~~\frac{\eta_{\mathcal{P}}(1)s}{\omega(1-2\varepsilon)(\alpha_1+\beta_1-1)},\\
&~~~~~~\frac{(1-\varepsilon)b-c+\eta_\mathcal{P}(1)s}{\omega[(1-\varepsilon)\alpha_1+\varepsilon(1-\beta_1)+\gamma_1-2(1-\delta_1)]},\\ &~~~~~~\frac{\varepsilon b-c+\eta_\mathcal{P}(1)s}{\omega[(1-\varepsilon)\alpha_0+\varepsilon(1-\beta_0)+\gamma_0-2(1-\delta_0)]}\Big\}
 \end{split}
 \end{aligned}
  \right.
  \end{equation}

By solving Eq.(\ref{problem2}) with fixed  $\alpha_\theta=\beta_\theta=\gamma_\theta=\delta_\theta=1, \forall \theta\in\Theta$, we can obtain necessary and sufficient conditions for the existence of a feasible solution for the design problem of Eq.(\ref{problem2}) as follows
\begin{equation} {}
  \label{solution2}
  \begin{split}
\omega \geq \max\Big\{\frac{(2-\varepsilon)s}{b(1-2\varepsilon)^2},\frac{2[(1-\varepsilon)b-c]+(2-\varepsilon)s}{b(2-\varepsilon)(1-2\varepsilon)}\Big\}
 \end{split}
  \end{equation}

\textcolor{black}{Note that there exists a sustainable two-sided rating protocol under a stricter recommended strategy $\pi_s$ if and only if $\max\Big\{\frac{(2-\varepsilon)s}{b(1-2\varepsilon)^2},\frac{2[(1-\varepsilon)b-c]+(2-\varepsilon)s}{b(2-\varepsilon)(1-2\varepsilon)}\Big\}<1$.}

Given $\alpha_\theta=\gamma_\theta=\delta_\theta=1, \forall \theta\in\Theta$, the design problem in Eq.(\ref{problem2}) w.r.t $\beta_0$ and $\beta_1$ can be rewritten as
\begin{equation} {}
  \label{problem20}
  \left\{
  \begin{aligned}
  \begin{split}
&\mathop{\min}\limits_{(\beta_0,\beta_1)}~\frac{1}{1+\frac{2-\varepsilon\beta_0}{\varepsilon\beta_1}}\\
&s.t.~\frac{(1-2\varepsilon)b}{2+\omega\varepsilon(\beta_1-\beta_0)}\geq \\
&~~~~~~\max\Big\{\frac{s(2-\varepsilon\beta_1)}{\omega\beta_0(1-2\varepsilon)(2+\varepsilon\beta_1-\varepsilon\beta_0)},\\
&~~~~~~~~~~~~~\frac{s(2-\varepsilon\beta_1)}{\omega\beta_1(1-2\varepsilon)(2+\varepsilon\beta_1-\varepsilon\beta_0)},\\
&~~~~~~~~~~~~~\frac{[(1-\varepsilon)b-c](2+\varepsilon\beta_1-\varepsilon\beta_0)+s(2-\varepsilon\beta_1)}{\omega(2+\varepsilon\beta_1-\varepsilon\beta_0)(2-\varepsilon\beta_1)},\\ &~~~~~~~~~~~~~\frac{(\varepsilon b-c)(2+\varepsilon\beta_1-\varepsilon\beta_0)+s(2-\varepsilon\beta_1)}{\omega(2+\varepsilon\beta_1-\varepsilon\beta_0)(2-\varepsilon\beta_0)}\Big\}
 \end{split}
 \end{aligned}
  \right.
  \end{equation}

An important observation is that Algorithm 1 can be efficiently used to handle the design problem of Eq.(\ref{problem20}) in a similar manner, which is omitted here.

\section{\textcolor{black}{EVALUATION RESULTS}}

\begin{figure*}
\centering
\begin{minipage}{0.245\textwidth}
  \rightline{\includegraphics[width=\textwidth]{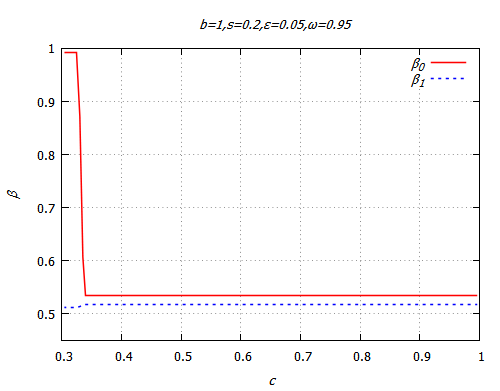}}
  \centerline{(\emph{a}) }
\end{minipage}
\begin{minipage}{0.245\textwidth}
  \leftline{\includegraphics[width=\textwidth]{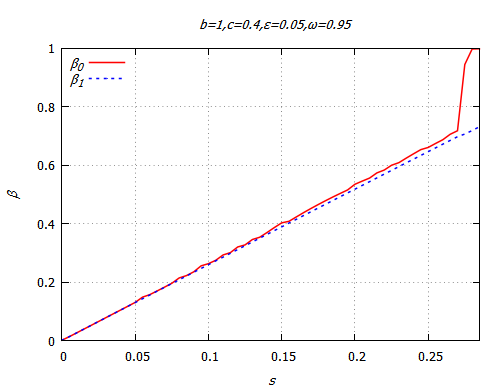}}
  \centerline{(\emph{b})}
\end{minipage}
\begin{minipage}{0.245\textwidth}
  \rightline{\includegraphics[width=\textwidth]{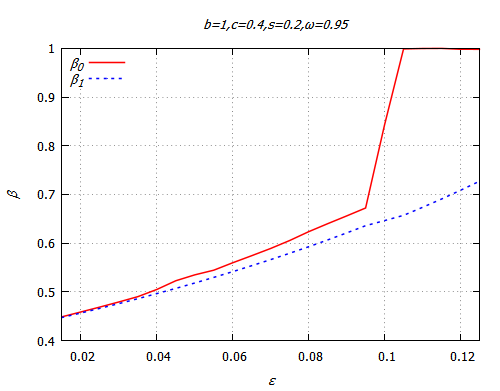}}
  \centerline{(\emph{c}) }
\end{minipage}
\begin{minipage}{0.245\textwidth}
  \leftline{\includegraphics[width=\textwidth]{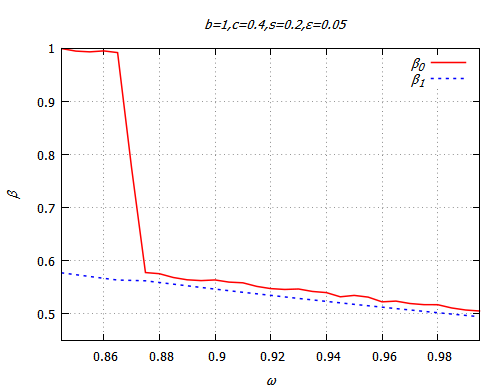}}
  \centerline{(\emph{d})}
\end{minipage}
\caption{The impact of design parameters $\beta_0$ and $\beta_1$ against intrinsic parameters: (a) $c$; (b) $s$; (c) $\varepsilon$; (d) $\omega$.}
\label{fig3}
\end{figure*}

\begin{figure*}
\centering
\begin{minipage}{0.245\textwidth}
  \rightline{\includegraphics[width=\textwidth]{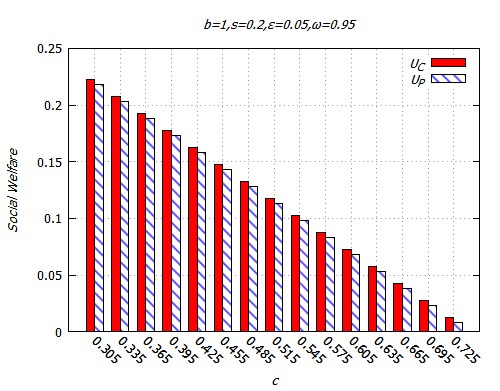}}
  \centerline{(\emph{a}) }
\end{minipage}
\begin{minipage}{0.245\textwidth}
  \leftline{\includegraphics[width=\textwidth]{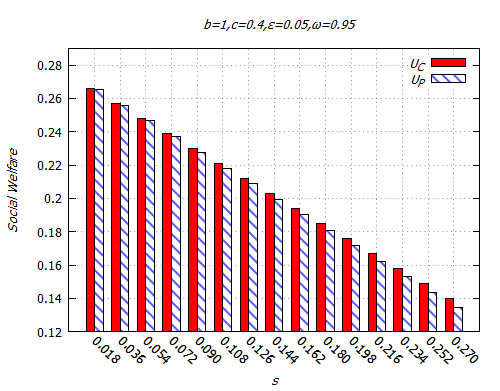}}
  \centerline{(\emph{b})}
\end{minipage}
\begin{minipage}{0.245\textwidth}
  \rightline{\includegraphics[width=\textwidth]{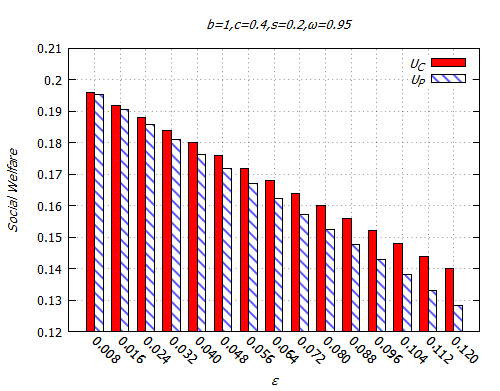}}
  \centerline{(\emph{c}) }
\end{minipage}
\begin{minipage}{0.245\textwidth}
  \leftline{\includegraphics[width=\textwidth]{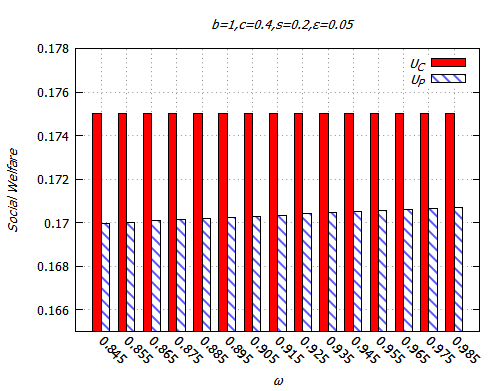}}
  \centerline{(\emph{d})}
\end{minipage}
\caption{Normalized performance against intrinsic parameters: (a) $c$; (b) $s$; (c) $\varepsilon$; (d) $\omega$.}
\label{fig4}
\end{figure*}

In this section, we provide numerical results to evaluate the key features of our proposed two-sided rating protocol designed for service exchange dilemma in crowdsourcing. First of all, we show how to determine the optimal recommended strategy as intrinsic parameters vary. Secondly, we investigate the impact of intrinsic parameters on design parameters. Thirdly, we examine the performance of the optimal design of two-sided rating protocols. Throughout our experiments, the benefit $b$ for unit service exchange is normalized to be 1, cost $c$ and $s$ are restricted to be smaller than $b$. \textcolor{black}{Finally, we investigate how to implement the proposed protocol in a crowdsourcing based service exchange system in details.}

\subsection{Optimal Recommended Strategy Against Intrinsic Parameters}

Figure 2 shows that the \textcolor{black}{optimal} recommended strategy is determined by intrinsic parameters $c$, $s$, $\varepsilon$ and $\omega$. When both of $c$ and $\omega$ are sufficiently large (given $b$=1, $s=0.2$ and $\varepsilon=0.05$) as shown in Figure 2(a), the recommended strategy $\pi$ can be sustained, and thus be selected to be the optimal choice among the other two candidate recommended strategies $\pi_s$ and $\pi_0$. \textcolor{black}{Under $\pi_0$, the server will be recommended to provide low-quality service regardless of his own and his client's ratings.} As $c$ or $\omega$ decreases, $\pi$ cannot be sustained any more, hence, the optimal recommended strategy changes from $\pi$ to $\pi_s$. The main reason behind this phenomenon is that, smaller $c$ and $\omega$ introduce a higher probability for self-interested users to deviate from the principle of fairness, which needs a stricter recommended strategy $\pi_s$ to increase users' incentive to comply with the social norm. When the region of $c$ and $\omega$ in which $\pi_s$ is not sustained, $\pi_0$ will be the unique sustainable recommended strategy, which yields zero social welfare for each user, since full cooperation cannot be achieved in such a scenario.

Figure 2(b) shows the optimal recommended strategy with $b=1$, $c=0.4$ and $\varepsilon=0.05$ as $s$ and $\omega$ vary. As $s$ increases, a user with a higher $\omega$ has higher probability to be a compliant user. Otherwise, $\pi$ will not be sustained, and thus the stricter recommended strategy $\pi_s$ will be selected to be optimal. When there does not exist any sustainable rating protocol in the region of $s$ and $\omega$, in order to obtain maximal social welfare, the optimal recommended strategy changes from $\pi$ to $\pi_s$ and eventually to $\pi_0$. A similar phenomenon can be found in Figure 2(c), which plots the change of optimal recommended strategy versus $\varepsilon$ and $\omega$ with given $b=1$, $c=0.4$ and $s=0.2$.

\subsection{The Impact of Intrinsic Parameters on Design Parameters}

Figure 3 plots how the optimal design ($\beta_0$ and $\beta_1$) is influenced by intrinsic parameters $c$, $s$, $\varepsilon$, and $\omega$. There does not exist a sustainable two-sided rating protocol with the recommended strategy $\pi$ when $c$ is sufficiently small, as shown in Figure 3(a). This is because users will choose to be a client with a higher probability $\lambda>\frac{1}{2}$, that is they will deviate from the principle of fairness when $c$ is small (\emph{e.g.}, $c<0.305000$). As $c$ increases, we can obtain a higher $\beta_0$ and a lower $\beta_1$, this is due to the fact that a higher $c$ introduces a higher $\eta_\mathcal{P}(0)$, and hence it needs a higher $\beta_0$ to punish non-compliant users. When $c$ is sufficiently large, $\beta_0$ and $\beta_1$ will not change, since a large $c$ do not affect on the sustainable of rating protocols, and thus it is no necessary to enhance the punishment factors.

Figure 3(b) is very similar with Figure 3(c), these two figures plot the impact of $s$ and $\varepsilon$ on the design parameters $\beta_0$ and $\beta_1$, respectively. As $s$ and $\varepsilon$ increase, it becomes more difficult to incentivize users to comply with the social norm, and punishment factors $\beta_0$ and $\beta_1$ should be increased to sustain a rating protocol. Different from Figure 3(b) and 3(c), punishment factors $\beta_0$ and $\beta_1$ decrease as $\omega$ increases, this is because it is easier to give incentives to comply with the social norm and punishments through the designed two-sided rating protocol. It can be observed that $\beta_0$ is always larger than $\beta_1$ in all of the four subfigures in Figure 3, this is due to the fact that $\eta_\mathcal{P}(0)<\eta_\mathcal{P}(1)$, in order to sustain the designed two-sided rating protocol with sufficient punishment, a larger $\beta_0$ and a lower $\beta_1$ will receive a better performance.

\subsection{Performance Efficiency}

Figure 4 examines the performance of the optimal design two-sided rating protocol against intrinsic parameters $c$, $s$, $\varepsilon$ and $\omega$ (denoted as $U_\mathcal{P}$). For the comparison, the social optimum $(1-\varepsilon)b-c-s$ (denoted as $U_\mathcal{C}$) is considered, which can be exactly achieved only by users with $\rho=1$ (or $\lambda=\frac{1}{2}$), who provide high-quality service all the time when they are matched as servers, however, it is not an equilibrium. In other words, such a social optimum is impossible to be achieved. Our goal is to be as close as possible to this social optimum.

The social welfare $U_\mathcal{P}$ monotonically decreases with $c$, $s$ and $\varepsilon$, but increases with $\omega$, where $U_\mathcal{C}$ is only determined by $c$, $s$ and $\varepsilon$, and is independent of $\omega$ with given $b=1$. In Figure 4(a), the performance gap between $U_\mathcal{P}$ and $U_\mathcal{C}$ is almost unchanged. The major reason is that the impact of $c$ on punishment factors can almost be ignored, thus both $U_\mathcal{P}$ and $U_\mathcal{C}$ change only against the value of $c$ with given $b$, $s$, $\varepsilon$ and $\omega$. While as shown in Figure 4(b) and 4(c), the performance gap between $U_\mathcal{P}$ and $U_\mathcal{C}$ becomes more significant as $s$ and $\varepsilon$ increase, respectively. This is because the incentive to comply with the social norm decreases and so is his one-period utility, and hence punishment factors $\beta_0$ and $\beta_1$ will be increased to provide sufficient incentive, thus reduces the social welfare $U_\mathcal{P}$. In particular, the gap between $U_\mathcal{P}$ and $U_\mathcal{C}$ is gradually narrowed as $\omega$ increases, as shown in Figure 4(d). Since users have more patience when the discount factor $\omega$ of users increases from 0 to 1, it becomes easier to sustain a two-sided rating protocol, thereby leading to a decrease in punishment factors and an increase in social welfare.

\subsection{\textcolor{black}{Implementation Issues}}
\textcolor{black}{As shown in Figure \ref{model}, we assume that the protocol designer stands at the crowdsourcing platform's point of view, and explores the strategy of users aiming to maximize their utilities. The implementation of the proposed protocol consists of five stages:}
\begin{enumerate}[(i)]
  \item \textcolor{black}{The protocol designer measures intrinsic parameters (\emph{i.e.}, $b$, $s$, $c$, $\varepsilon$ and $\omega$) of the system.}
  \item \textcolor{black}{According to part A of Section \uppercase\expandafter{\romannumeral5}, the protocol designer investigates whether there exists a sustainable two-sided rating protocol, as well as how to choose the optimal recommended strategy, both of which are determined by intrinsic parameters.}
  \item \textcolor{black}{The optimal design parameters (\emph{i.e.}, $\alpha_\theta$, $\beta_\theta$, $\gamma_\theta$, $\delta_\theta$, $\forall\theta\in\Theta$) can be calculated from the results in part B of Section \uppercase\expandafter{\romannumeral5}.}
  \item \textcolor{black}{Based on the calculated result in stage (\romannumeral3), the performance of the proposed protocol can be examined.}
  \item \textcolor{black}{By comparing the gap between the theoretical and practical values of social welfare, the protocol designer updates the design parameters in order to increase the practical value of social welfare.}
\end{enumerate}

\textcolor{black}{We divide the above five stages into two parts: (1) Direct implementation of stages (\romannumeral1), (\romannumeral2), (\romannumeral3) and (\romannumeral4) is a one-round implementation, which is based on evaluation results obtained in the first half of Section \uppercase\expandafter{\romannumeral5}. (2) Bargaining implementation of stage (\romannumeral5) in a trial-and-error way. This is because users may not be entirely rational in a real crowdsourcing scenario, and hence the protocol designer should decrease reward factors ($\alpha_\theta$, $\gamma_\theta$, $\forall\theta\in\Theta$) and increase punishment factors ($\beta_\theta$, $\delta_\theta$, $\forall\theta\in\Theta$) to compel users to contribute well behaviors in the initial stage, and then try to update reward factors as well as punishment factors gradually in order to be closer to the theoretical value. However, technical details on how to update the design parameters are beyond the scope of this work and are left for future study.}

\section{CONCLUSION AND FUTURE WORK}

In this paper, we proposed a service exchange dilemma in a two stage game, and developed a game-theoretic design of two-sided rating protocol to stimulate cooperation among self-interested users, and thus overcome the inefficiency of the socially undesirable equilibrium. By rigorously analyzing how users' behaviors are influenced by intrinsic parameters, design parameters, as well as users' valuation of their individual long-term utilities, we characterize the optimal design by selecting eight optimal design parameters ($\alpha_\theta$, $\beta_\theta$, $\gamma_\theta$, $\delta_\theta$), $\forall\theta\in\Theta$, where we proved that $\alpha_\theta^\ast=\gamma_\theta^\ast=\delta_\theta^\ast=1, \forall\theta\in\Theta$ is always the optimal solution of Eq.(\ref{problem}), and designed a two-stage two-step algorithm to select $\delta_\theta^\ast,\forall\theta\in\Theta$ which achieves low-complexity computation in an alternate manner. The social welfare $U_\mathcal{P}$ obtained by our proposed two-sided rating protocol $\mathcal{P}$ can be very close to the social optimum, especially when users are sufficiently patient and the monitoring and
reporting error is small.

\textcolor{black}{Although we have proved that the service exchange dilemma can be efficiently solved by the proposed two-sided rating protocol, one might be curious about the effect of implementing more elaborate two-sided rating protocols. In the following, we point out a few extendable directions for future work.}

\textcolor{black}{First, rating labels can be extended from two levels to multiple levels. As shown in our previous work \cite{6}, to study multi-level problem, one should consider how to determine an optimal size $K$ of rating labels, as well as a threshold value $\kappa$ that a user with $\theta\ge \kappa$ or $\theta<\kappa$ will be rewarded or punished, respectively. In this case, two additional design parameters are added to the two-sided rating protocol design problem, which greatly increases the computation complex of the problem. How to find out an algorithm with a lower computation complexity is a challenging problem.}

\textcolor{black}{Second, it is interesting to design multiple levels of actions for a server in the second stage game. In order to incentivize self-interested users to contribute good behaviors, an elaborate rating update rule $\tau$ consisting of more reward factors and punishment factors should be carefully designed.}

\textcolor{black}{Third, taking into account continuous rating labels is a challenging task, because reward factors and punishment factors should be designed as convex functions and concave functions, respectively, which greatly increases the computational complexity of optimal design.}

\textcolor{black}{Fourth, considering continuous actions for a server is another challenging task, as the design parameters will be replaced by elaborate continuous function.}

\textcolor{black}{Fifth, the above four cases can be combined pair by pair, which is an interesting problem. On the other hand, the difficulty of the problem is greatly enhanced, and is deferred to a future study.}

\ifCLASSOPTIONcompsoc
  \section*{Acknowledgments}
\else
  \section*{Acknowledgment}
\fi

This work was supported in part by the National Natural Science Foundation of China under Grant 61402418, 11531011, 11771013, 61503342, 61672468, 61602418, and in part by the Social Development Project of Zhejiang Provincial Public Technology Research under Grant 2017C33054.

\section*{\textcolor{black}{Appendix}}
\subsection{\textcolor{black}{Computation Process for Table \ref{tab02}}}
\begin{table}[tb]

\caption{Payoff matrix of each user for the second-stage under the (C,S) case}
\label{tab03}
\begin{center}
\begin{tabular}{c|c|c|c|c|c|c|}

  \multicolumn{1}{c}{\textbf{}}& \multicolumn{6}{c}{\textbf{$user 2$}} \\
  \multicolumn{1}{c}{\textbf{}}& \multicolumn{3}{c}{\textbf{$H$}}& \multicolumn{3}{c}{\textbf{$L$}} \\
\cline{2-7}
  \multicolumn{1}{c}{\textbf{$user 1$}}& \multicolumn{3}{|c|}{\quad$(1-\varepsilon)b-c$, $-s$\quad\quad}& \multicolumn{3}{c|}{\quad\quad$\varepsilon b-c$, $0$\quad\quad\quad} \\
\cline{2-7}
\end{tabular}%
\end{center}
\end{table}

\textcolor{black}{The computation process for Table \ref{tab02} in four cases is shown as follows:}

\textcolor{black}{Case $\uppercase\expandafter{\romannumeral 1}$: ($C,S$), \emph{i.e.}, user 1 requests services as a client, and user 2 chooses to be a server.}

\textcolor{black}{When a user requests services as a client, a matching rule is used to determine corresponding server. We model the interaction between a pair of matched users in the second stage as a gift-giving game \cite{gift}, and the payoff matrix of the gift-giving game between a client and a server is presented in Table \ref{tab03}. We assume that $b>\frac{c+s}{1-\varepsilon}$ so that the service of a user creates a positive net social benefit, and social welfare is maximized when all servers choose action $H$ in the gift-giving games they play, which yields payoff $(1-\varepsilon)b-c-s$ to every user. On the contrary, action $L$ is the dominant strategy for the server, which constitutes a Nash equilibrium of the gift-giving game.}

\textcolor{black}{We now take a step back and compute expected utilities for such a case. When the client consumes a cost $c$ for choosing $C$ in the first stage, and receives $\varepsilon b-c$ payoff in the second stage, the server will choose $L$ and suffer a low cost, which is approximated by 0 here. Such a case also results in zero payoff. We summarize this in the $CS$ cell of the pay-off matrix in Table \ref{tab02}.}

\textcolor{black}{Case $\uppercase\expandafter{\romannumeral 2}$: ($S,C$), \emph{i.e.}, user 1 chooses to be a server and user 2 chooses to be a client. Case $\uppercase\expandafter{\romannumeral 2}$ is symmetric with case $\uppercase\expandafter{\romannumeral 1}$. Thus, the ex-ante utility of user 1 is $v_1=0$, and the ex-ante utility of user 2 is $v_2=\varepsilon b-c$.}

\textcolor{black}{Case $\uppercase\expandafter{\romannumeral 3}$: ($C,C$), \emph{i,e.}, both users choose to request services as clients. Each user consumes a cost $c$, but receives zero service benefit as there is no user offering service. We describe this in the $CC$ cell of the pay-off matrix in Table \ref{tab02}.}

\textcolor{black}{Case $\uppercase\expandafter{\romannumeral 4}$: ($S,S$), \emph{i.e.}, The expected utility of each user is zero as no user requests service. The $SS$ cell of the pay-off matrix in Table \ref{tab02} describes such a case.}

\subsection{\textcolor{black}{Computation Process for Eq.(\ref{bp20})}}
\textcolor{black}{Given a rating protocol $\mathcal{P}$ and a chosen rate $\lambda$, we can derive the expected payoff of a $\theta$-user according to Eq.(\ref{pp}) as follows:}
\begin{equation} {}
  \label{eq43}
  \left\{
  \begin{aligned}
  \begin{split}
  \textcolor{black}{v_{\mathcal{P},\lambda}(0)=}&\textcolor{black}{\lambda[{\eta(0)b_{\pi}(0, 0)}+{\eta(1)b_{\pi}(1, 0)}]}\\
  &\textcolor{black}{-(1-\lambda)[{\eta(0)c_{\pi}(0,0)}+{\eta(1)c_{\pi}(0,1)}]}\\
  \textcolor{black}{v_{\mathcal{P},\lambda}(1)=}&\textcolor{black}{\lambda[{\eta(0)b_{\pi}(0, 1)}+{\eta(1)b_{\pi}(1, 1)}]}\\
  &\textcolor{black}{-(1-\lambda)[{\eta(0)c_{\pi}(1,0)}+{\eta(1)c_{\pi}(1,1)}]}
 \end{split}
 \end{aligned}
 \right.
 \end{equation}

 \textcolor{black}{On the one hand, a user chooses $C$ and $S$ with probabilities $\lambda$ and $1-\lambda$, respectively. And whether a user is matched with a 0-user or a 1-user is determined by $\eta_\mathcal{P}(0)$ and $\eta_\mathcal{P}(1)$, respectively. On the other hand, a user can receive service benefits by choosing $C$ and consume a cost to provide service by choosing $S$. As $b_\pi(0,0)=b-c$, $b_\pi(1,0)=0$, $c_\pi(0,0)=s$, and $c_\pi(0,1)=s$, and taking the monitoring error $\varepsilon$ into consideration, $v_{\mathcal{P},\lambda}(0)$ in Eq.(\ref{eq43}) can be rewritten as}
\begin{equation}
\label{eq44}
\begin{aligned}
\textcolor{black}{v_{\mathcal{P},\lambda}(0)=\lambda\big\{[\eta_\mathcal{P}(0)(1-\varepsilon)+\eta_\mathcal{P}(1)\varepsilon]b-c\big\}-(1-\lambda)s}
\end{aligned}
\end{equation}

\textcolor{black}{Similarly, as $b_\pi(0,1)=b-c$,\ $b_\pi(1,1)=b-c$,\ $c_\pi(1,0)=0$,\ $c_\pi(1,1)=s$, and given $\varepsilon$, $v_{\mathcal{P},\lambda}(1)$ in Eq.(\ref{eq43}) can be rewritten as}
\begin{equation}
\label{eq45}
\begin{aligned}
\textcolor{black}{v_{\mathcal{P},\lambda}(1)=}&\textcolor{black}{\lambda[(1-\varepsilon)b-c]-(1-\lambda)\eta_\mathcal{P}(1)s}
\end{aligned}
\end{equation}

 \textcolor{black}{Given a rating protocol $\mathcal{P}$ and a chosen rate $\lambda$, the expected long-term utilities of a 0-user and a 1-user according to Eq.(\ref{lpv}) can be derived as follows:}
 \begin{equation} {}
  \label{eq46}
  \left\{
  \begin{aligned}
  \begin{split}
  \textcolor{black}{v_{\mathcal{P},\lambda}^\infty (0)=} &\textcolor{black}{v_{\mathcal{P},\lambda} (0)+ \omega [p_{\mathcal{P},\lambda} (0 |0)v_{\mathcal{P},\lambda}^\infty (0)}\\
  &\textcolor{black}{+p_{\mathcal{P},\lambda} (1 |0)v_{\mathcal{P},\lambda}^\infty (1)]}\\
  \textcolor{black}{v_{\mathcal{P},\lambda}^\infty (1)= }&\textcolor{black}{v_{\mathcal{P},\lambda} (0)+ \omega [p_{\mathcal{P},\lambda} (0 |1)v_{\mathcal{P},\lambda}^\infty (0)}\\
  &\textcolor{black}{+p_{\mathcal{P},\lambda} (1 |1)v_{\mathcal{P},\lambda}^\infty (1)]}
 \end{split}
 \end{aligned}
 \right.
 \end{equation}

\textcolor{black}{By substituting Eq.(\ref{eq44}) and Eq.(\ref{eq45}) into the RHS of Eq.(\ref{eq46}), we have}
\begin{equation}
\label{eq47}
\begin{aligned}
\textcolor{black}{v_\mathcal{P}^\infty (1)-v_\mathcal{P}^\infty (0)=\frac{\lambda\eta_\mathcal{P}(1)(1-2\varepsilon)b+(1-\lambda)\eta_\mathcal{P}(0)s}{1+\omega(\phi_1-\phi_0)}}
\end{aligned}
\end{equation}

\ifCLASSOPTIONcaptionsoff
  \newpage
\fi



%



%

\end{document}